\documentclass[
 reprint,
 amsmath,amssymb,
 aps,
]{revtex4-2}

\usepackage{dcolumn}
\usepackage{bm}

\usepackage{siunitx}
\usepackage[italic]{hepnames}
\usepackage[super]{nth}

\usepackage{graphicx}
\usepackage{natbib}

\begin{document}

\title{
Absolutely Scintillating:\\
constraining $\nu$ mass with black hole-forming supernovae}

\author{George Parker}
\email{geparker@uni-mainz.de}
\author{Michael Wurm}
\email{michael.wurm@uni-mainz.de}
\affiliation{
 PRISMA+ Cluster of Excellence and Institut für Physik, \\
 Johannes Gutenberg-Universität Mainz, 55099 Mainz, Germany\\
}

\date{\today}

\begin{abstract}
The terrestrial detection of a neutrino burst from the next galactic core-collapse supernova (CCSN) will provide profound insight into stellar astrophysics, as well as fundamental neutrino physics. Using Time-Of-Flight (ToF) effects, a CCSN signal can be used to constrain the absolute neutrino mass. In this work, we study the case where a black hole forms during core-collapse, abruptly truncating the neutrino signal. This sharp cutoff is a feature that can be leveraged in a ToF study, enabling strict limits to be set on the neutrino mass which are largely model-independent. If supernova neutrinos are detected on Earth in liquid scintillator detectors, the exceptional energy resolution would allow an energy-dependent sampling of the ToF effects at low neutrino energies. One promising experimental program is the Jiangmen Underground Neutrino Observatory (JUNO), a next-generation liquid scintillator detector currently under construction in China. Using three-dimensional black hole-forming core-collapse supernova simulations, the sensitivity of a JUNO-like detector to the absolute neutrino mass is conservatively estimated to be $m_\nu < 0.39^{+0.06}_{-0.01}$ \unit{\electronvolt} for a 95\% CL bound. A future-generation liquid scintillator observatory like THEIA-100 could even achieve sub-0.2 \unit{\electronvolt} sensitivity.

\end{abstract}

\maketitle

\section{Introduction}

Since massless neutrinos were refuted experimentally by the discovery of neutrino flavour oscillations \cite{Fukuda1998-jq, Ahmad2002-xy}, the neutrino absolute mass scale has been an open question. Neutrinos are less than $\sim10^{-6}$ the mass of the next lightest elementary particle. A non-zero neutrino mass is physics Beyond-the-Standard Model, and the relative size of this mass suggests a novel process for mass generation. Therefore, measuring the exact mass of the neutrino is of great scientific interest and has been a subject of intense experimental and theoretical work.\par

Massive neutrinos from core-collapse supernovae (CCSNe) propagate across an extreme distance before reaching Earth, embedding an energy- and mass-dependent delay in the signal. This Time-Of-Flight (ToF) effect can be used to set an upper limit on the absolute neutrino mass.
The idea of using neutrinos from core-collapse supernovae (CCSNe) as a probe of neutrino mass was first described by Zatsepin \cite{Zatsepin:1968kt}. This effect is the delay of a massive neutrino compared to a massless particle travelling at the speed of light. It is derived from the delay time between two ultrarelativistic particles (for the general formulation see \cite{Stodolsky2000-ab,Fanizza2016-id,Fleury2016-lj}). At lowest order, the relation between the delay $\Delta t$, distance $D$, neutrino mass $m_\nu$, and neutrino energy $E_\nu$, is:

\begin{equation}
\label{eq:1}
\Delta t = \frac{D}{2c} \left( \frac{m_\nu}{E_\nu}\right)^2,
\end{equation}

\noindent which can be rewritten in convenient units, where the D is in units of $10$ kpc, which the average distance to the Galactic centre \cite{Beacom2000-gp, Beacom2001-qf}:
\begin{equation}
\label{eq:2}
\Delta t [\text{s}] = 0.515 \left( \frac{m_\nu [\text{eV}]}{E_\nu [\text{MeV}]} \right)^2 D [\text{10 kpc}].
\end{equation}

When neutrinos from core-collapse supernova SN1987A were detected on Eart, it presented the first, and only, opportunity to apply ToF techniques. Kamiokande II detected 12 events \cite{Hirata1987-pw}, IMB detected 8  \cite{Bionta1987-so}, and 5 events were detected in Baksan \cite{Alexeyev1988-ts}. Using the data from Kamiokande II and IMB, an upper limit derived of ${m_\nu} < 12.0$ \unit{\electronvolt} (90\% CL) \cite{Arnett1987-zg}. Later, this limit was tightened using a parameterized model of CCSN neutrino emission and better analysis of backgrounds (allowing the use of the Baksan events) achieving a limit of ${m_\nu} < 5.7$  \unit{\electronvolt} \cite{Loredo2002-fr}. This was updated to ${m_\nu} < 5.8$ \unit{\electronvolt} \cite{Pagliaroli2010-do} with an improved parameterization and likelihood analysis. \par

Beacom \textit{et al} \cite{Beacom2000-gp, Beacom2001-qf} first pointed out that if a CCSN evolves into a black hole, a much tighter neutrino mass limit can be set using the ToF effect. Black hole formation creates an abrupt cutoff in the neutrino signal, making ToF effects clearer, and leading to an improved mass sensitivity compared to a different stellar evolution. Instead of carrying out a fit on the entire neutrino curve \cite{Pagliaroli2010-do}, black hole formation allows the consideration of events only immediately before and after the black hole formation time, $t_{\textrm{BH}}$. This limits the systematic uncertainty introduced by the largely experimentally undetermined model for CCSN neutrino emission. Beacom \textit{et al} considered a neutrino signal with black hole formation in the context of the Super-Kamiokande detector \cite{Abe2016-zo}, deriving a sensitivity of 1.8 \unit{\electronvolt} for a canonical CCSN at 10 kpc distance \cite{Beacom2000-gp, Beacom2001-qf}. \par 

Lu \textit{et al} \cite{Lu2015-uu} explored the neutrino mass constraint from CCSN neutrinos detected with a next-generation liquid-scintillator detector. In particular, they demonstrate that in the Jiangmen Underground Neutrino Observatory (JUNO) \cite{An2016-kb} it would be possible to set a mass limit of $m_\nu <$ 0.83 $\pm$ 0.24 \unit{\electronvolt} (95\% CL), competitive with current laboratory limits. However, this bound is obtained without black hole formation. In this work, we show that if the signal of a black hole-forming core-collapse supernova is observed in a liquid scintillator detector, the low detection threshold and excellent energy resolution would allow for a fine-grained analysis of the potential ToF signature, allowing a model-independent and competitive constraint on the absolute neutrino mass to be achieved. \par

The present paper is structured as follows: Sec. \ref{sec:level1} is a short review of the fundamentals of neutrino mass and existing measurements. Sec. \ref{sec:level2} is a brief explanation of CCSNe evolution with the corresponding structure of the neutrino emission emphasised in Sec. \ref{sec:level3}. Sec. \ref{sec:level4} is a discussion of the target channel (inverse beta decay), with a short overview of other next-generation neutrino detectors for similar studies. A description of the simulations and parameterizations used in this study follows in Sec. \ref{sec:level5}. Sec. \ref{sec:level6} develops the approach used to arrive at an upper $m_{\nu}$ limit that could be set in case of non-observation. The resulting neutrino mass sensitivity is presented in Sec. \ref{sec:level7}, with the systematic uncertainties of this technique evaluated in Sec. \ref{sec:level8}. The effects of the detector mass and distance to Earth are quantified in Sec. \ref{sec:level9} and \ref{sec:level10}, respectively. Sec. \ref{sec:level11} is devoted to systematic effects that have the potential to weaken the derived mass limit, but in practice do not have a large influence on the derived bound.

\section{Background}
\subsection{\label{sec:level1}{Neutrino Mass Measurements}}

Neutrino flavour oscillations are dependent on the mass-squared differences, ${\Delta}m_{ij}^2={\Delta}m_{i}^2-{\Delta}m_{j}^2$. However, the sign of the mass difference ${\Delta}m_{31}$ is not known, leaving two possible neutrino mass patterns in nature. The first is 'Normal Ordering' (NO), where $m_3 > m_2 > m_1$, and the second is 'Inverted Ordering' (IO), where $m_2 > m_1 > m_3$ \cite{Scholberg2018-ms}. 

Although oscillation phenomena are not sensitive to the absolute scale of neutrino masses, we can set lower limits on the sum of neutrino mass states, $\sum {m_\nu}$, using the mass-squared differences. If we set the lightest mass state to zero, $m_0 = 0$, in Eq. (\ref{eq:3}) [NO] and Eq. (\ref{eq:4}) [IO], we calculate limits of $\sum {m_\nu} \gtrsim 0.06$ \unit{\electronvolt} and $\sum {m_\nu} \gtrsim 0.1$ \unit{\electronvolt}, respectively \cite{Choudhury2018-jh}.\par
\begin{equation}
\label{eq:3}
\sum {m_\nu} = m_0 + \sqrt{m_0^2 + {\Delta}m_{21}^2} 
+ \sqrt{m_0^2 + {\Delta}m_{31}^2}
\end{equation}
\begin{equation}
\label{eq:4}
\sum {m_\nu} = m_0 + \sqrt{m_0^2 + {\Delta}m_{31}^2} 
+ \sqrt{m_0^2 + {\Delta}m_{31}^2 + {\Delta}m_{21}^2}
\end{equation}

Cosmological probes including the Cosmic Microwave Background, Type Ia supernovae, and Baryon Acoustic Oscillation observations are also sensitive to the sum of the neutrino mass states, giving an extremely strong upper limit, $\sum {m_\nu} < 0.11$ \unit{\electronvolt} (95\% CL) \cite{Planck_Collaboration2020-vu}. This limit can be improved to  $\sum {m_\nu} < 0.09$ \unit{\electronvolt} (95\% CL) in the most constraining case \cite{Di_Valentino2015-he}, putting it in a weak tension with the lower limit for IO. However, these limits are strongly dependent on cosmological models and therefore other measurements of the absolute neutrino mass are needed to corroborate this value \cite{Planck_Collaboration2020-vu, Di_Valentino2015-he}.\par

The strongest limit on the neutrino mass from direct measurements is the results from the KATRIN experiment, which uses the kinematics of tritium $\beta$-decays to derive an upper mass limit for $\bar{\nu}_e$ of 0.8 \unit{\electronvolt} (90\% CL) \cite{The_KATRIN_Collaboration2022-jx}. KATRIN is projected to be able to achieve a mass limit approaching 0.2 \unit{\electronvolt} (90\% CL) across its full experimental run \cite{KATRIN:2005fny}. 

The absolute neutrino mass limit could be further advanced by the Project 8 experiment, an advanced experimental program also targeting tritium $\beta$-decays with a new spectroscopy method based on cyclotron radiation. The projected final sensitivity of Project 8 is 0.04 eV \cite{Esfahani2017-og}.

\subsection{\label{sec:level2}{Black Hole-Forming Core-Collapse Supernovae}}

When a massive star (mass $\gtrsim 8M_{\odot}$, where $M_{\odot}$ is the solar mass) exhausts its nuclear fuel, its core collapses producing a neutron star or black hole. Considering historical supernovae, the rate of Galactic supernova can be calculated to be
$R=3.2^{+7.3}_{-2.6}$ (century)${}^{-1}$ \cite{Adams2013-yj}, but could also be as low as $R=1.63 \pm 0.46$ (century)${}^{-1}$ from a conservative, combined analysis \cite{Rozwadowska2021-vc}.\par

The fraction of massive stars that end their life as a black hole compared to a neutron star is known as $f_{\textrm{BH}}$. This quantity is not theoretically well-known, with estimates in the region $f_{\textrm{BH}} \lesssim 0.30–0.35$ \cite{OConnor2011-vl}. Observational surveys with the Large Binocular Telescope give a value of 
$f_{\textrm{BH}}=0.16^{+0.23}_{-0.12}$ (90\% CL) with 11 years of data \cite{Neustadt2021-zj}.

In this section, we will briefly describe the stages of core-collapse, and then go into detail on the resulting neutrino flux (for detailed reviews of the CCSN evolution, see \cite{Janka2012-ds, Burrows2013-ca, Janka2017-wn}).\par

\paragraph{Pre-Explosion}
The chain of nuclear fusion has resulted in an iron core surrounded by layers of successively lighter elements ('onion-shell' structure) \cite{Janka2012-ds, Burrows2013-ca, Janka2017-wn}.

\paragraph{Core-collapse and -bounce}
The outer core collapses supersonically, and the inner collapses subsonically \cite{Janka2017-wn}. As the core density increases, neutrinos become trapped. The inner core rebounds against the infalling outer core \cite{Burrows2013-ca}, and a shock wave forms at the interface.

\paragraph{Shock propagation and stall}

The shock propagates through the surrounding material \cite{Janka2017-wn}. Electron neutrinos produced by electron capture on free protons, $\Pelectron + p \rightarrow \Pnue + n$, are suddenly free, causing the 'neutronization' burst in the $\Pnue$ neutrino signal \cite{Janka2012-ds, Burrows2013-ca, Janka2017-wn}. The shock continues to lose energy with it eventually stalling \cite{Janka2017-wn, Li2021-no, Walk2020-pj}.

\paragraph{Accretion and neutrino heating}

The dense core reaches approximate hydrostatic equilibrium with the surroundings \cite{Li2021-no} with matter accreting onto the shock front \cite{Janka2012-ds}. Large-scale, non-radial instabilities such as standing accretion-shock instability (SASI) can develop \cite{Blondin2003-sq}, where the shock front undergoes a violent sloshing and spiral oscillation, mixing the material underneath and leading to mass asymmetry \cite{Janka2017-wn, Walk2020-pj}. 

\paragraph{Black hole formation}
In the scenario relevant for this work, the explosion is not successful, and matter continues to accrete onto the 'proto-neutron star' until it collapses into a black hole. \par

\subsection{\label{sec:level3}{Expected Neutrino Signal}}

\begin{figure}[h]
\hspace{-1.1cm}
\includegraphics[width=265pt]{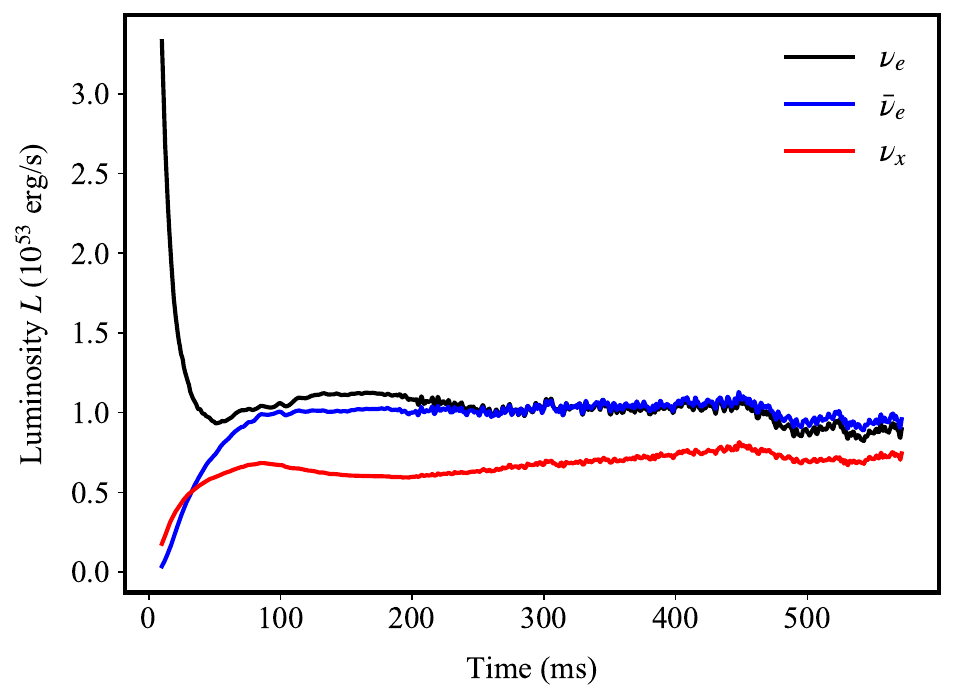}
\caption{\label{fig:fig1} The neutrino luminosity for $\Pnue$ (black), $\bar{\nu}_e$ (blue), and $\nu_x$ (red). Where $v_x$ represents the other flavours ($\Pnum$, $\bar{\nu}_\mu$, $\Pnut$, $\bar{\nu}_\tau$) which are predicted to act almost identically in the context of core-collapse. Model (s40) from \cite{Walk2020-pj}, accessed through the Garching CCSN Archive \cite{Garching}.}
\end{figure}

The expected neutrino emission begins at core-bounce, with 'time' in the CCSN simulations meaning the post-bounce time, $t_{\textrm{pb}}$ \cite{Li2021-no}. The sharp peak in $\Pnue$ luminosity in Fig. 1 is the neutronization burst. Next, we expect a large, quasistationary flux in the accretion phase, with the potential for SASI to cause noticeable modulations in the flux corresponding to the SASI frequency. Finally, black hole formation during the accretion phase abruptly truncates the neutrino signal at $t_{\textrm{BH}}$ \cite{Li2021-no, Walk2020-pj}. Black hole formation, the neutronization burst, and even SASI modulations could be effective targets for ToF analyses. \par

If no black hole formed, a successful explosion would be followed by a cooling phase (see \cite{Janka2017-wn}). The cooling phase would constitute the majority of the $\nu$ signal \cite{Li2021-no} in this case. 

\subsection{\label{sec:level4}{Detection for ToF studies}}

In liquid scintillator and water detectors, electron antineutrinos, $\bar{\nu}_e$, participate in inverse beta decay (IBD) on free protons,
\begin{equation*}
\bar{\nu}_e + p \rightarrow \APelectron + n,
\end{equation*}

which is the dominant detection channel. It provides both the largest statistics ($\mathcal{O}(10^3)$) for next-generation liquid scintillator and water Cherenkov detectors \cite{Abe2016-zo} for a canonical CCSN at 10 kpc. Crucially for this study, IBD provides a direct relation between the neutrino and positron energy of the prompt signal. The delayed capture of the neutron on H (or Gd \cite{Ikeda2021-ar}) can be used to remove potential backgrounds from elastic neutrino-electron scattering and other interaction channels. For a review of the various channels and detectors for CCSN neutrinos, see \cite{Li2021-no}.

Super-K (Super-Kamiokande) and its planned successor experiment Hyper-K (Hyper-Kamiokande) \cite{Proto-Collaboration_Hyper-Kamiokande2018-zc} are Cherenkov detectors with fiducial volumes of 22.5 and 187 \unit{\kilo\tonne} of ultrapure water, respectively. Beacom \textit{et al} derive an upper limit of $m_\nu < 1.8 \unit{\electronvolt}$ considering the IBD channel in the Super-K detector \cite{Beacom2000-gp, Beacom2001-qf}. As the mass limit scales with detector mass as $m_\textrm{lim} \sim M_D^{-1/2}$, we comment that the result of 1.8 eV in Super-K from Beacom \textit{et al} can be trivially scaled for Hyper-K--giving a mass limit of $\sim 0.6$ eV for a detector $\sim8$ times larger.

In the present study, we focus on JUNO, the Jiangmen Underground Neutrino Observatory, a liquid scintillator detector situated southwest of Kaiping city in Guangdong province in southern China \cite{An2016-kb}. JUNO has a fiducial mass of 20 \unit{\kilo\tonne} and has 17,612 20-inch PMTs. The detector is currently under construction and is projected to begin data taking in 2024 \cite{JUNO2022}. 

While JUNO will provide similar statistics in the IBD channel to Super-K, JUNO has a better energy resolution $\delta E/E \sim 3\% / \sqrt{E(\unit{\mega\electronvolt})}$ and an exceptional energy threshold of 0.2 \unit{\mega\electronvolt} \cite{An2016-kb}. 
As the ToF effect, $\Delta t \propto \frac{1}{E^2}$ (Eq. 2), is most pronounced at lower energies, we expect JUNO to be substantially more sensitive to neutrino mass than Super-K.

We also consider THEIA-100, a proposed 100 \unit{\kilo\tonne} detector containing water-based liquid scintillator (WbLS) \cite{Askins2020}. By doping ultrapure water with liquid scintillator, it is possible to achieve a balance between Cherenkov light which allows impressive directional resolution, and scintillation light which gives excellent energy resolution. 
In \cite{Zsoldos}, it has been shown that WbLS with 5\% organic loading would be sufficient to achieve an MeV-level threshold. Given the large target mass of THEIA-100, we expect an extremely tight limit to be set on the neutrino mass, exceeding the projected final sensitivity of KATRIN.

\section{Methods}
\subsection{\label{sec:level5}{Simulated Supernova Models}}

The sharp cutoff of the neutrino signal by black hole formation is largely model-independent and therefore provides an excellent feature to search for a mass-induced ToF delay. We use three-dimensional black hole-forming CCSN simulations to generate realistic IBD spectra and statistics, but the general analysis approach does not depend on the models used. 

The two non-rotating black hole-forming models s40 \cite{Woosley2007-nz} and u75 \cite{Woosley2002-xx} originate from progenitors of mass 40$M_{\odot}$ and 75$M_{\odot}$, respectively. According to the common notation of CCSN simulations, the prefix 's' denotes solar metallicity ($Z_{\odot} \sim 0.0134$) and the prefix 'u' denotes ultra-low metallicity ($Z \sim 10^{- 4} Z_{\odot}$) (additionally the prefix 'z' denotes zero metallicity) \cite{OConnor2011-vl}.

The models s40 and u75 are realised in three-dimensional hydrodynamical simulations with three neutrino flavours ($\Pnue$, $\bar{\nu}_e$, and $\nu_x$) \cite{Walk2020-pj}. The simulations were carried out with \texttt{PROMETHEUS-VERTEX} code \cite{Rampp2002-qw} which approximates general relativity effects with a modified Tolman-Oppenheimer-Volkoff potential \cite{Marek2005-if}. Further details of the simulation can be found in \cite{Walk2020-pj}.\par

Access to these models is possible with the Garching CCSN Archive \cite{Garching}. The neutrino data is given in the parameters of neutrino luminosity, $L$, mean energy $\langle E \rangle$, and second moment of energy,  $\langle E^2 \rangle$, for each time-step and flavour. The pinching factor, $\alpha$, which is a measure of the spectral pinching (with $\alpha = 2$ being a Maxwell-Boltzmann spectrum), can be calculated using \cite{Keil2003-kf}:
\begin{equation}
\frac{\langle E^2 \rangle}{\langle E \rangle^2} = \frac{2+\alpha}{1+\alpha}.
\label{eq:5}
\end{equation}

The (normalized) pinched-thermal functional form \cite{Tamborra2012-vu, Scholberg2018-ms} can be used to construct the neutrino flux information, $\phi(E)$, for each flavour:
\begin{eqnarray}
\Phi^0(E) & = \left(\frac{L}{\langle E\rangle}\right)\frac{(\alpha+1)^{(\alpha+1)}}{\Gamma(\alpha+1)}\left(\frac{E}{\langle E \rangle}\right)^\alpha \exp \left(-\frac{(\alpha+1)E}{\langle E \rangle}\right).
\label{eq:6}
\end{eqnarray}

The Mikheyev-Smirnov-Wolfenstein (MSW) effect occurs at the outer envelope, due to the supernova density profile and resonant flavour conversions \cite{Wolfenstein1978-wy, Wolfenstein1979-cc, Dighe2000-rv, Takahashi2001-ah, Kawagoe2010-hh}. Assuming adiabatic transitions, this can be written in the form \cite{Lu2016-cv} for NO:
\begin{subequations}
\begin{eqnarray}
    \Phi_{\Pnue} = & \Phi^0_{\Pnue},\\
    \Phi_{\bar{\nu}_e} = & \Phi^0_{\bar{\nu}_e}{\cos^2\theta_{12} + \Phi^0_{\nu_x}\sin^2\theta_{12}}, \\
    \Phi_{\nu_x} = & \Phi^0_{\nu_x}\frac{2+\cos^2\theta_{12}}{4}
    + \Phi^0_{\Pnue}\frac{1}{4}
    + \Phi^0_{\bar{\nu}_e}\frac{\sin^2\theta_{12}}{4}
    \label{eq:7}
\end{eqnarray}
\end{subequations}

and for IO we have:
\begin{subequations}
\begin{eqnarray}
    \Phi_{\Pnue} =& \Phi^0_{\Pnue}{\sin^2\theta_{12} + \Phi^0_{\nu_x}\cos^2\theta_{12}}, \\
    \Phi_{\bar{\nu}_e} =& \Phi^0_{\nu_x}, \\
    \Phi_{\nu_x} =& \Phi^0_{\nu_x}\frac{2+\sin^2\theta_{12}}{4}
    + \Phi^0_{\bar{\nu}_e}\frac{1}{4}
    + \Phi^0_{\Pnue}\frac{\cos^2\theta_{12}}{4},
    \label{eq:8}
\end{eqnarray}
\end{subequations}

where $\theta_{12}$ is the solar mixing angle, which is taken to be $33.45^{+0.77}_{-0.75}$$^{\circ}$ \cite{Esteban2020-ha}.\par 

The event rate is calculated for the canonical distance to a galactic supernova, $D = 10$ kpc. For JUNO, we use the number of free protons in the detector, $N_p = 1.5 \times 10^{33}$, and the IBD cross-section, $\sigma_{\textrm{IBD}}$ \cite{Strumia2003-dx}.
\begin{equation}
    \dfrac{\mathrm{d}^2 N}{\mathrm{d}E\mathrm{d}t} = N_p \frac{1}{4\pi D^2}\Phi_{\bar{\nu}_e} \sigma_{\textrm{IBD}},
    \label{eq:9}
\end{equation}

\begin{figure}[b!]
\hspace{-0.74cm}
\includegraphics[width=260pt]{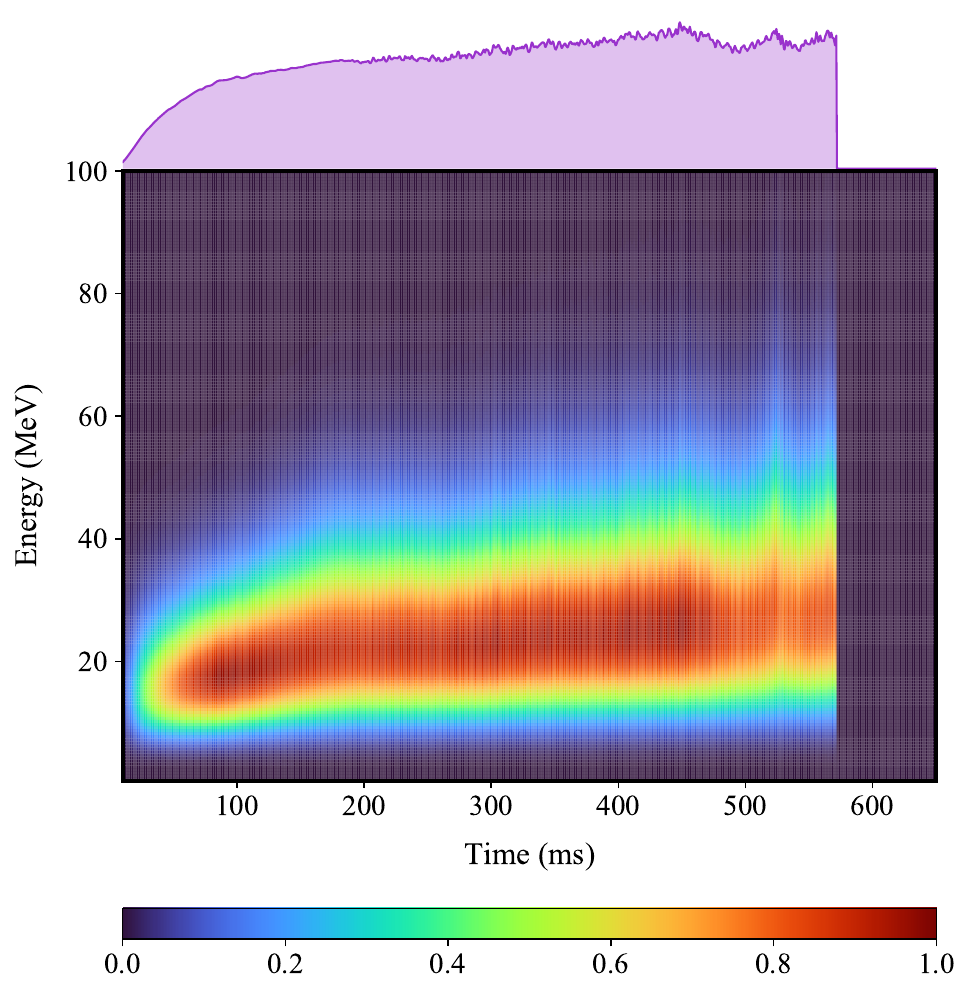}
\caption{\label{fig:fig2} Detected IBD ($\bar{\nu}_e$) spectrum in JUNO as a function of energy and time, with colours representing normalized number of neutrino events. A projection of the energy-integrated event profile is included. Black hole formation with a massless neutrino ensures no ToF delay, producing the sharp cutoff of events at $t_{\mathrm{BH}} \sim$ 570 ms. See text for more information.}
\end{figure}

where input variables and smearing effects have been dropped for simplicity. The nominal JUNO energy resolution and threshold have been used when calculating energy smearing \cite{An2016-kb}. Note that in this case energy is the detected positron energy $E_{\APelectron} = E_{\nu} - 1.293$ \unit{\mega\electronvolt}. 

For s40 we expect $\sim$ 7000 IBD events in JUNO, with $\sim$ 4000 events for u75. Although s40 has a lower flux rate at $t_{\mathrm{BH}}$ compared to u75, the integrated event number is higher because the black hole forms at a later time ($\sim$ 570 ms post-bounce for s40, and $\sim$ 250 ms post-bounce for u75).

As part of this study, we used the supernova neutrino package \texttt{SNOwGLoBES} \cite{Scholberg-2021}, which includes multiple detector configurations and detection channels and can be used to approximate the fluxes, mixing effects, and expected event rates.

Fig. 2 is a 2D plot of the detected IBD ($\bar{\nu}_e$) event spectrum in JUNO to highlight the energy-time structure of the signal. A projection is given at the top of the plot to show the classic energy-integrated event profile. 
For all of the following plots we use model s40 from \cite{Walk2020-pj} accessed from the Garching CCSN Archive \cite{Garching}, with the canonical distance from Earth (10 kpc) and normal neutrino mass ordering (NO), unless stated otherwise.

\subsection{\label{sec:level6}{Statistical Procedure}}

In this paper, we use the numerical method in \cite{Lu2015-uu} as a guide. In this work, Lu \textit{et al} also constrain the absolute neutrino mass using CCSN neutrinos detected in JUNO, but without black hole formation. Their method involves a parameterization of neutrino emission \cite{Pagliaroli2009-ci} with eight parameters: one for the absolute start time, one for the early or 'rising' time, three for the accretion phase, and three for the cooling phase. The method in \cite{Lu2015-uu} fits all eight parameters to artificial data with varying neutrino mass delays, and performs a likelihood analysis to derive an upper limit of the absolute neutrino mass. This provides a model-dependent bound of 0.83 $\pm$ 0.24 \unit{\electronvolt} (95\% CL) for NO.

However, black hole formation allows a great simplification of this analysis, and ensures the mass limit is largely independent of the supernova model. Since the black hole forms during the accretion phase, the neutrino flux and spectrum can be assumed as quasistationary for a time period of several 10 ms before $t_{\mathrm{BH}}$. In our analysis, we use a period of 75 ms before black hole formation to extract the spectral shape and rate of neutrino signal prior to its cutoff. This fitted region is presented in Fig. 3. While our fit procedure assumes the fit to be stationary, flux and energy are expected to vary considerably to effects such as SASI oscillations. Our fit window is chosen sufficiently long to average out these effects. 

\begin{figure}[t!]
\hspace{-0.74cm}
\includegraphics[width=260pt]{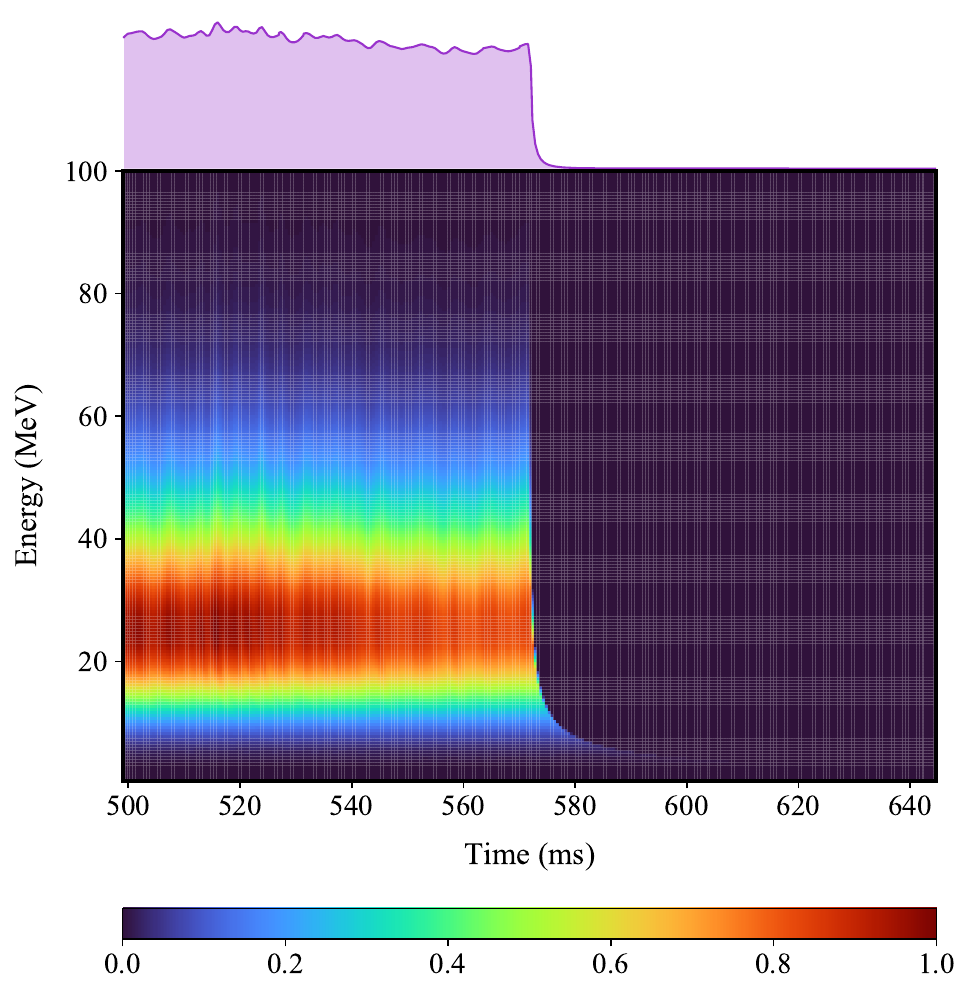}
\caption{\label{fig:fig3} Fitted region of detected IBD ($\bar{\nu}_e$) spectrum in JUNO as a function of energy and time, with colours representing normalized number of neutrino events. A projection of the energy-integrated event profile is included. Black hole formation and ToF delay corresponds to a 1.0 eV neutrino mass. See text for more information.
}
\end{figure}

For this period, we create artificial data sets with a uniform spectra using the pinched-thermal functional form in Eq. (6). We then induce a ToF delay in the artificial data sets consistent with absolute neutrino masses from $0.0-1.5$ \unit{\electronvolt} in steps of $0.025$ \unit{\electronvolt}. A binned likelihood function with Poisson statistics is used to compare the data from the simulation and the artificial data set to derive a median sensitivity. For each neutrino mass step, we calculate the $\Delta\chi^2$ with 4 free parameters: the time of black hole formation $t_{\textrm{BH}}$ (to mitigate the effect of $t_{\textrm{BH}}$ being wrongly characterised), the pinching parameter, $\alpha$, the mean energy, $\langle E \rangle$, and the normalisation, $\mathcal{N}$. 

\section{Results}

\subsection{\label{sec:level7}{Constraint}}

To derive limits for the $m_\nu$ sensitivity, we apply the technique outlined in Sec. \ref{sec:level6} to the model signal spectra discussed in Sec. \ref{sec:level5}. For this, we investigate two progenitor models (s40 and u75), both possible neutrino mass orderings in nature (NO and IO), as well as different assumptions for visibility of SASI in the signal. We present both a mass limit from where the neutrino emission has been spherically averaged (Fig. 2 and 3), and from the specific observer direction corresponding to the highest number of events (denoted 'Direction 1' for s40 and the 'Strong Modulations' direction for u75). The  specific observer direction and spherically averaged cases are compared in Fig. 4 for model s40. A 2D energy-time plot of neutrino events for the observer direction is presented in Fig. 5 to be contrasted with the spherically averaged case in Fig. 2.

\hspace{-4cm}
\begin{figure}[t!]
\hspace{-0.75cm}
\includegraphics[width=265pt]{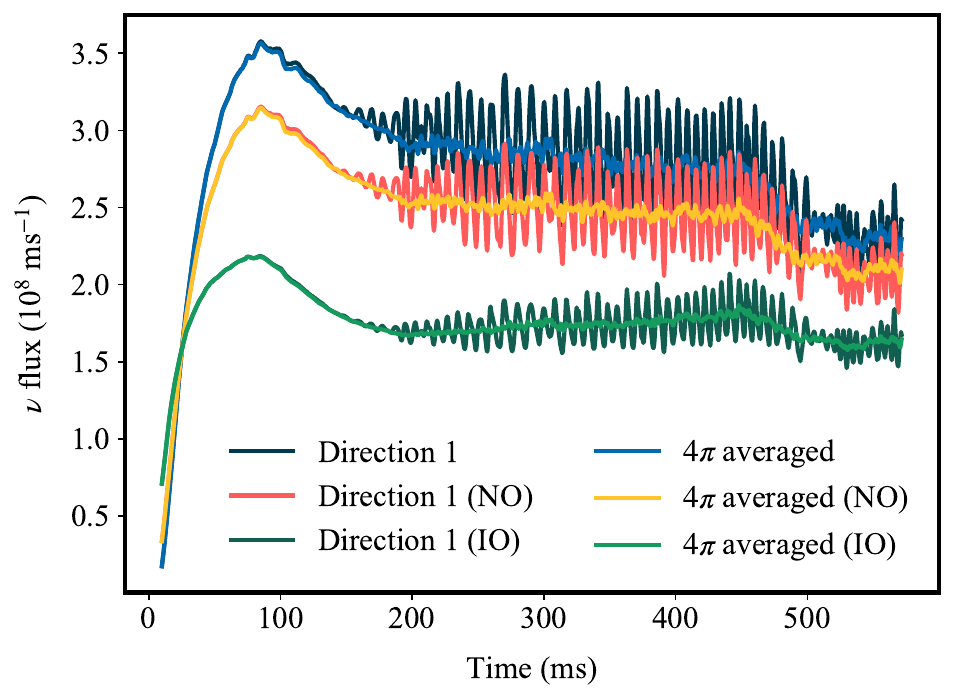}
\caption{\label{fig:fig4} Electron antineutrino ($\bar{\nu}_e$) flux for 'Direction 1' and the spherically averaged case for 1) no mixing, 2) normal ordering (NO) and 3) inverted ordering (IO). Model s40 \cite{Walk2020-pj} accessed from \cite{Garching}.
}
\end{figure}

The signals from the specific observer directions have the imprint of violent SASI activity. By also obtaining a mass limit from these signals we highlight the robustness of this method. As our constant fit averages over several oscillations we expect that SASI will not influence our neutrino mass sensitivity.

We try different SASI realizations but always obtain very similar mass limits because our constant fit averages over several oscillations. Hence, SASI is not expected to affect our neutrino mass sensitivity.

Fig. 4 also shows the effect of the adiabatic MSW effect for NO and IO. This effect lowers the $\bar{\nu}_e$ flux, with the IO having the larger decrease when compared to NO. The plot of luminosities (Fig. 1) shows in the accretion phase $\Phi^0_{\Pnue} \approx \Phi^0_{\bar{\nu}_e} > \Phi^0_{x}$. Equations (7b) and (8b) show that $\bar{\nu}_e$ is a full swap with $\nu_x$ for IO, and a superposition of $\Pnue$ and $\nu_x$ for NO, explaining the discrepancy in Fig. 4. This reduction in the total IBD events for IO results in a less restrictive mass limit compared to NO. Conversely, for $\Pnue$ in DUNE for example, the relevant equations would be (7a) and (8a), and IO would ensure a tighter constraint than NO.

\begin{table}[h]
\caption{\label{tab:table1}
Calculated upper limit for $m_\nu$ at 95\% CL for the two spherically averaged models and possible mass orderings. All values in \unit{\electronvolt}.}
\begin{ruledtabular}
\begin{tabular}{ccdd}
Mass Ordering&40$M_\odot$ model&
\multicolumn{1}{c}{\textrm{75$M_\odot$ model}}\\
\hline \\ [-1.5ex]
NO&0.40&0.32\\
IO&0.45&0.35\\
\end{tabular}
\end{ruledtabular}
\end{table}

Exact upper limits in Table I are derived using the method described in the previous section (in Sec. IIIB); producing a $\Delta\chi^2$ curve and then requiring $\Delta\chi^2 > 3.84$ for 95\% CL, using the $\Delta\chi^2$ distribution with one degree of freedom. It should be noted we do not follow the more complex procedure from \cite{Lu2015-uu} where the deviation from the true $\Delta\chi^2$ distribution is investigated which means that we are placing a slightly less stringent limit as would be the case in the more accurate treatment.

\begin{figure}[t!]
\hspace{-0.74cm}
\includegraphics[width=260pt]{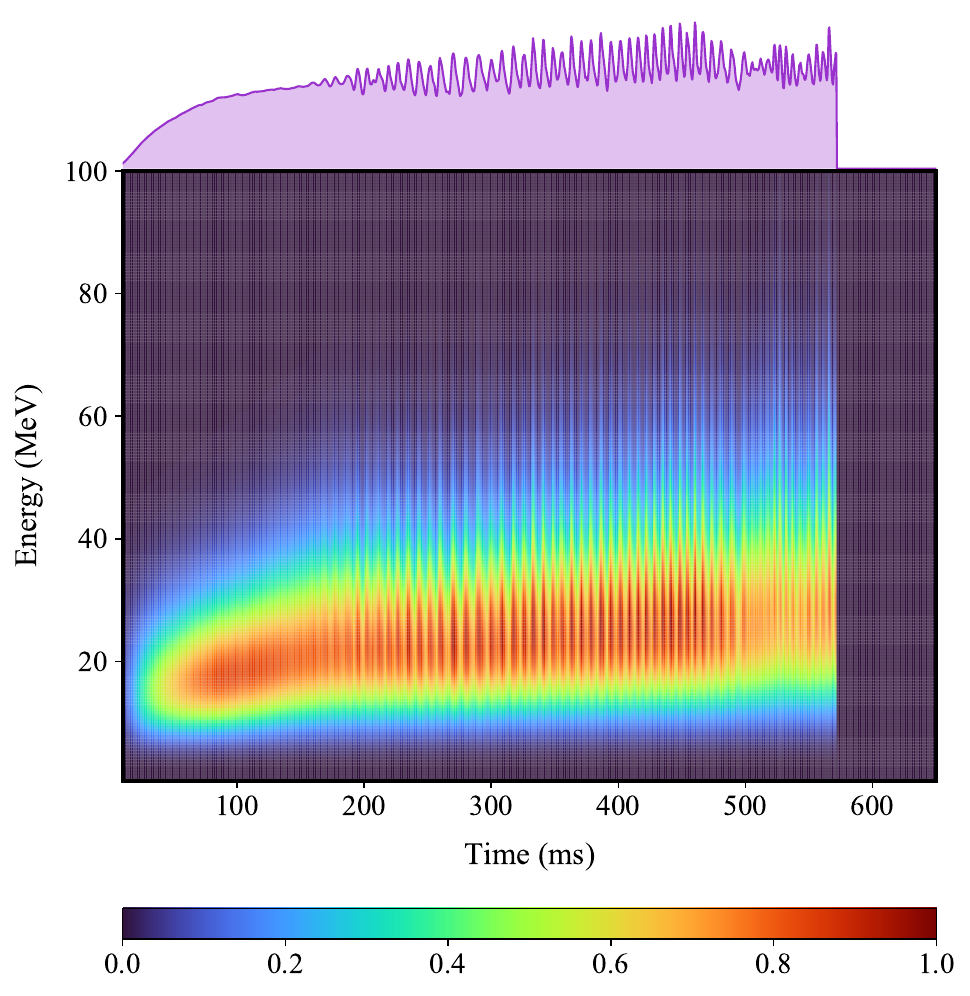}
\caption{\label{fig:fig5} Plot of Direction 1 to be compared with the spherically averaged plot (Fig. 2). Detected IBD ($\bar{\nu}_e$) spectrum in JUNO as a function of energy and time, with a projection of events in time. Normalized neutrino events are represented by colours. Black hole formation with a massless neutrino ensures no ToF delay.}
\end{figure}

Model u75 provides a tighter constraint of 0.32 eV compared to 0.40 eV for s40 (for NO), as it has a higher event rate in the detector immediately before black hole formation. Note that progenitor mass and neutrino luminosity are not related in a simple way. The neutrino output is dependent on many interrelated factors including metallicity, stellar structure, mass accretion rate, and other complexities in the stellar evolution. These factors and their effect on the post-bounce dynamics can be encapsulated by a parameter called 'compactness' \cite{OConnor2011-vl} which has been shown to correlate with neutrino emission in 1D \cite{OConnor2013-at}.

\begin{table}[h]
\caption{\label{tab:table2}%
Relative differences in $m_\nu$ upper limit for the specific direction compared to the spherically averaged case at 95\% CL. }
\begin{ruledtabular}
\begin{tabular}{ccdd}
Mass Ordering&40$M_\odot$ model&
\multicolumn{1}{c}{\textrm{75$M_\odot$ model}}\\
\hline \\ [-1.5ex]
NO&-2.9\%&0.1\%\\
IO&-1.5\%&0.3\%
\end{tabular}
\end{ruledtabular}
\end{table}

Table II shows the effect of modulations on the derived neutrino mass limit. The model s40 has a larger relative difference in mass bound between modulated and average signal of -2.9\% compared to 0.1\% for u75 (for NO). We expect this result as the model s40 has a significantly stronger SASI oscillation in the signal. For s40, the SASI oscillation lowers the limit, and for u75, modulations in the signal increase the mass bound, but it is clear the magnitude of this effect on the derived limit is small. 

\subsection{\label{sec:level8}{Statistical Uncertainties from Bootstrapping}}

\begin{figure}[t!]
\hspace{-1.1cm}
\includegraphics[width=265pt]{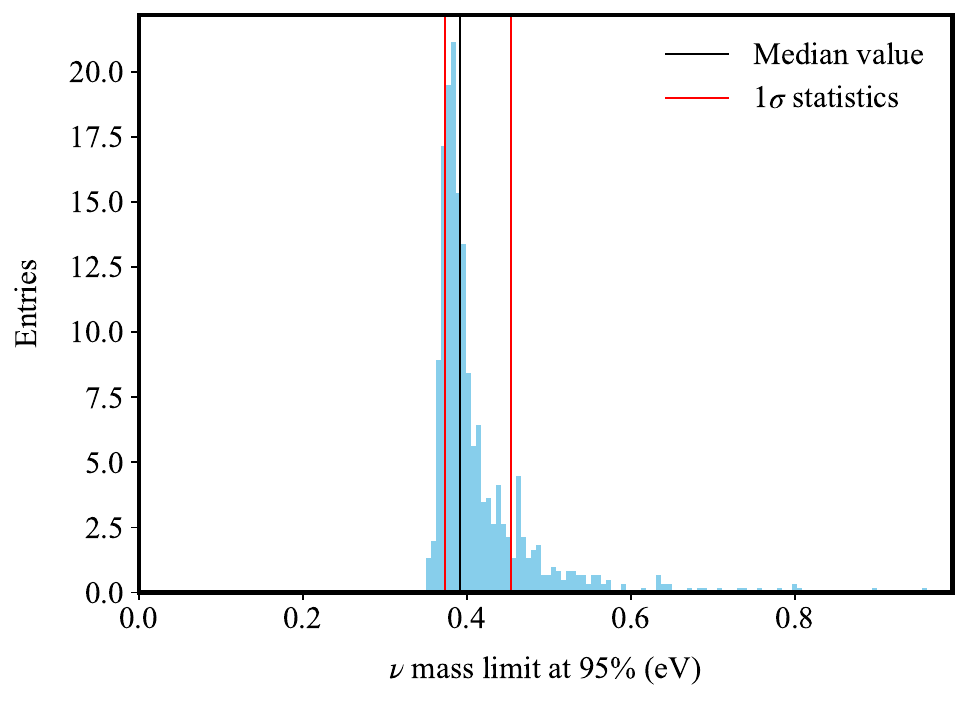}
\caption{\label{fig:fig6} Histogram of the mass sensitivity at 95\% CL with 1000 pseudo-experiments simulated using the s40 model with NO as the representative 'Asimov' dataset \cite{Isaac, Cowan2011-iv}.}
\end{figure}

The exact limit on $m_\nu$ will be dependent on the statistical fluctuations in the signal. Here, we quantify the level to which the upper limit on $m_\nu$ is affected by these fluctuations by performing a statistical bootstrap. The spherically averaged s40 model with NO is chosen as a representative case. The event rate for the model is used as a 'true' distribution from which 1000 simulated pseudo-experiments (toy MC) are produced. A histogram is generated from the calculated neutrino mass limit for each pseudo-experiment (Fig. 6). The median value of this dataset is 0.39 eV, which is approximately equal to the bound of 0.40 eV derived in the previous section. Uncertainty can then be estimated by considering 1$\sigma$ statistics around the median, leading to a result of $0.39^{+0.06}_{-0.01}$ eV for a 95\% CL bound (see Fig. 6). This distribution is clearly asymmetric with a maximum of 0.96 \unit{\electronvolt}. Therefore, the uncertainty is also asymmetric, showing that statistical fluctuations could have a large effect on the derived limit.

\subsection{\label{sec:level9}{Detector Mass}}

\begin{figure}[t!]
\hspace{-0.6cm}
\includegraphics[width=261pt]{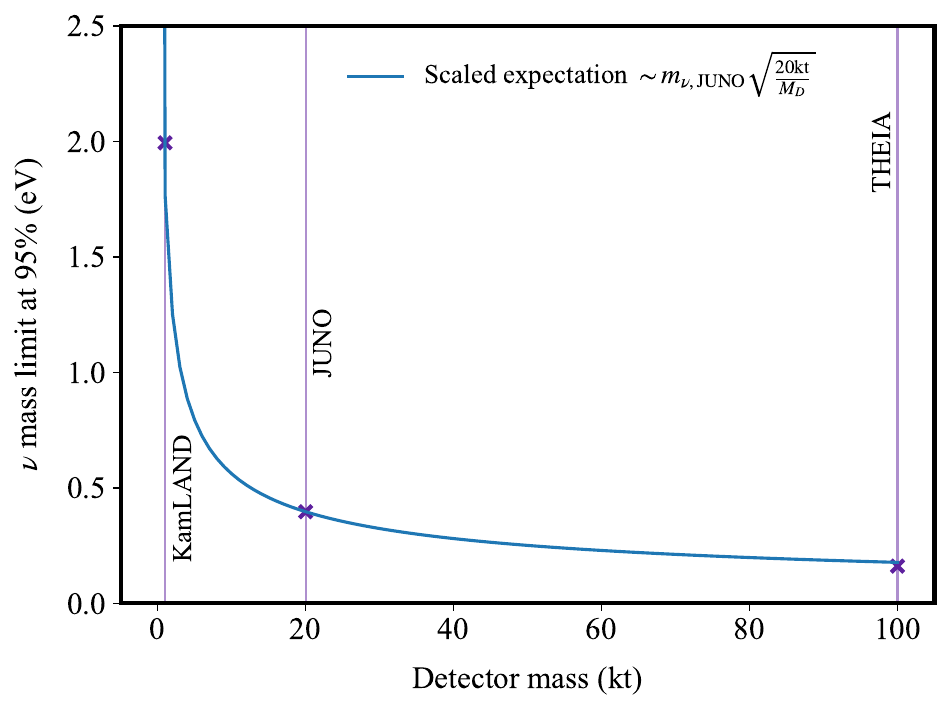}
\caption{\label{fig:fig7} The mass sensitivity of the black hole-forming CCSN at 95\% CL is shown for a range of detector masses. This range is chosen to include KamLAND, JUNO and THEIA(-100). The theoretical expectation scaled from the JUNO limit with Eq. (10) is a good fit for the other detectors.}
\end{figure}

In the rest of this work, we consider neutrinos being detected in the liquid scintillator detector JUNO, which currently under construction in southern China and is the most promising LS observatory in the imminent future. However, there are many other liquid scintillator detectors that would be suitable for a study of this type, from the past, the present and from proposals, such as Borexino \cite{Alimonti2009}, SNO+ \cite{Albanese2021}, and LENA \cite{Wurm2012}. 

Along with JUNO, we select two representative detector setups to cover a range of relevant liquid scintillator target masses. We select a 1 \unit{\kilo\tonne} detector setup which we label "KamLAND" and corresponds to the capabilities of KamLAND-Zen \cite{KamLAND, Abe2023} and SNO+. We also consider a 100 \unit{\kilo\tonne} setup which we call "THEIA" corresponding to THEIA-100 \cite{Askins2020}. In our setup, where we do not consider realistic detector efficiencies and response, the effect of different energy resolutions was found to have a negligible impact the $\nu$ mass limit derived. Therefore, we use the JUNO value of $3\% / \sqrt{E(\unit{\mega\electronvolt})}$ instead of $6.5\% / \sqrt{E(\unit{\mega\electronvolt})}$ and $\sim 7\% / \sqrt{E(\unit{\mega\electronvolt})}$ for KamLAND \cite{Abe2008} and THEIA, respectively.
For CCSN without black hole formation, the mass limit scales with detector mass as $m_\textrm{lim} \sim M_D^{-1/4}$. With the sharp cutoff induced by the formation of a black hole, Beacom \textit{et al} \cite{Beacom2000-gp, Beacom2001-qf} show that the derived mass limit instead goes with
\begin{equation}
    m_{lim} \sim \left(L_{BH} M_D\right)^{-1/2}.
    \label{eq:10}
\end{equation}
In Fig. 7, we show the KamLAND mass limit to be $1.99$ eV for a 95\% CL bound compared to a limit of $0.40$ eV for JUNO. To directly compare this to the limit from Beacom $\textit{et al}$, we have to account for both detector mass, $M_D$, and neutrino luminosity per flavour at the cutoff, $L_{BH}$ (Eq. 10). In \cite{Beacom2000-gp, Beacom2001-qf}, the $L_{BH}$ is $10^{52}$ erg/s, which is $\sim 1$ order of magnitude smaller than the neutrino luminosity in Fig. 1. Therefore, the limit derived by Beacom $\textit{et al}$ can be rescaled to $0.57$ eV showing that the analysis technique enabled by liquid scintillator detectors is a $\sim 30\%$ improvement. For THEIA, we derive a neutrino mass constraint of $0.16$ eV, which is below the projected final sensitivity for the KATRIN experiment of $0.2$ eV.

\subsection{\label{sec:level10}{Distance from Earth}}

In the earlier sections, we assume the canonical distance of a CCSN from Earth of 10 kpc, which is slightly farther than the average distance to the galactic centre \cite{Hunt2016-ua}. However, it is possible that the next detected CCSN could occur anywhere in the Galaxy and its satellites. SN1987A was determined to be in the Large Magellanic Cloud (LMC), $\sim 50$ kpc from Earth \cite{Pagliaroli2009-ci}. In the simulation paper for the s40 and u75 models \cite{Walk2020-pj}, it is suggested that the neutrino signal from these models would even be detectable at a few sigma significance from near-neighbour galaxies \cite{Walk2020-pj}. However, it is not expected that there would be mass sensitivity at this distance due to a lack of statistics.\par

In this section, we quantify the effect of distance on the neutrino mass constraint. The distance, $D$, has two effects on the neutrino emisssion: 1) the ToF delay scales with $\propto D$, and 2) the flux rate scales with $\propto \frac{1}{D^2}$. \par

For CCSN without black hole-formation, the mass limit is independent of the distance from Earth, $D$. With the sharp cutoff, Beacom \textit{et al} \cite{Beacom2000-gp, Beacom2001-qf} show that the derived mass limit instead goes with
\begin{equation}
    m_{lim} \sim D^{1/2}.
    \label{eq:11}
\end{equation}

\begin{figure}[b!]
\hspace{-1.1cm}
\includegraphics[width=260pt]{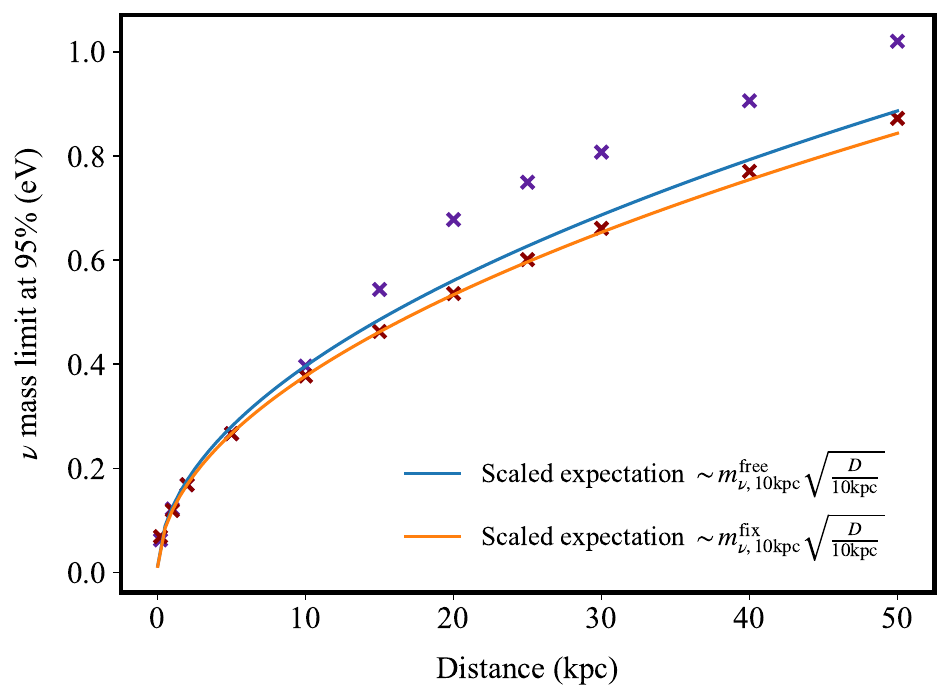}
\caption{\label{fig:fig8} The mass sensitivity of the black hole-forming CCSN at 95\% CL is shown from 0.2-50 kpc. This range is chosen because it is the distance from Betelgeuse to the LMC, where it is thought SN1987A originated. The purple crosses denote the classic scenario with four free parameters, (formation time, $t_{\textrm{BH}}$; pinching parameter, $\alpha$; mean energy, $\langle E \rangle$; and the normalisation, $\mathcal{N}$). The maroon crosses are generated by fixing the black hole formation time, and leaving the other three parameters free. The theoretical expectations scaled from the 10 kpc limit in each scenario are the blue and orange lines.}
\end{figure}

The distance from Earth in Fig. 8 ranges from 0.2 to 50 kpc. The distance to SN1987A is 50 kpc, where mass sensitivity at 95\% CL is $\sim 1$ \unit{\electronvolt}, larger than the current direct experimental limit of 0.8 \unit{\electronvolt}. The distribution of Galactic supernovae is within a narrow band, with 10\% of SNe within 5 kpc and 90\% within 15 kpc \cite{Adams2013-qa}. A limit rivalling KATRIN's projected sensitivity of of 0.2 \unit{\electronvolt} (90\% CL) \cite{KATRIN:2005fny} could be obtained for our CCSN model (s40) less than $\sim 5$ kpc from Earth. 

Although there are several CCSN candidates within 0.1-1 kpc, these events are rare. There are at least six supernova candidate stars $<0.3$ kpc from Earth, including Betelgeuse in the Orion constellation at a distance of $\sim 0.2$ kpc \cite{Firestone2014-oi}. In Fig. 8, we show that for our model (s40) located $0.2$ kpc from Earth,  we derive a mass limit of 0.06 eV. However, for very near-Earth supernova, there is a concern about the possibility of data loss as this extremely large rate could overload the data-acquisition systems \cite{Li2021-no}. Note that the data-acquisition system of JUNO will be designed to accommodate this possibility.

The orange curve in Fig. 8 is the relationship between the distance from Earth and neutrino mass limit expected from Eq. (11). The neutrino mass limits derived for different distances (the maroon crosses) follow this trend if the black hole-formation time is fixed. However, if the black hole-formation time allowed to be a free parameter, as in the rest of this work, we find significant deviation from the theoretical expectation (blue curve) and the mass limits (purple crosses). This is because the relationship between the cutoff time, normalisation, and distance from Earth is not captured by Eq. (11). This also exemplifies the importance of properly identifying the black hole formation time, which relies on the laudable timing resolution of LS detectors.

\section{\label{sec:level11}{Discussion}}
During the analysis, we have neglected certain 'higher-order' effects that could potentially alter the mass sensitivity. In this section, we attempt to quantify these effects, including Earth-matter effects, non-radial neutrino emission, and neutrino echoes.\newline

\paragraph{Earth-matter effects}

 As neutrinos travel through the Earth on their journey to a detector, they undergo the MSW effect. In this work, we took the most optimistic case, where the CCSN is in the same hemisphere as JUNO, and Earth-matter effects are not relevant. However, if Earth-matter effects are relevant they would cause a slight oscillation in the $\bar{\nu}_e$ energy spectrum for NO and would have no effect for IO \cite{Scholberg2018-ms, Borriello2012-me}. Of all current or near-future detectors, it is proposed that JUNO, as a liquid scintillator detector with excellent energy resolution, is best poised to disentangle this effect \cite{Liao2016-eo}. We can safely neglect this effect, as although it may cause a small oscillation in the neutrino spectrum, we have shown this technique is robust against even violent SASI oscillations. \newline

\paragraph{Rotation and non-radial neutrino emission}
In this work, we have only considered non-rotating models. It is conjectured that rotation would alter the flavour conversion and broaden the peaks of SASI activity \cite{Walk2018-bk}, although it is not expected that rotation would have a large effect on the mass sensitivity alone, it is important to consider rotation in combination with non-radial neutrino emission. 
In a simple picture, where neutrinos are emitted only radially from a non-rotating CCSN that undergoes black hole formation, the neutrinos either would not escape the expanding event horizon or continue on their radial trajectories. However, in a more realistic model neutrinos will be emitted in all directions at $t_{\textrm{BH}}$. As a consequence, some neutrinos will be trapped spiralling out in unstable circular orbits before escaping the gravitational field. The additional path length in case of this non-radial emission could soften the neutrino cutoff and therefore, introduce a tail of events similar to the mass-induced ToF effect. For non-rotating systems, the number of trailing events after the cutoff can be estimated to 0.7 events in JUNO, which is negligible compared to the ToF effect \cite{Wang2021-og}. However, for strongly rotating models, it is thought that this effect is much more important, and altering the size and the shape of the tail after black hole formation \cite{Wang2021-og}. Although the strongly rotating case is thought to be rare, this effect makes it important to characterize the energy-dependence of the cutoff, leveraging the unique low-energy signature of ToF delay. \newline

\paragraph{Neutrino Echoes}
Neutrino echoes are neutrinos emitted just before black hole formation which scatter on infalling material on their journey to Earth, thus being delayed \cite{Gullin2022-ez}. If this effect is present on the scale predicted in \cite{Gullin2022-ez} then in JUNO we expect $\sim5$ events after $t_{\textrm{BH}}$ from neutrino echoes. However, as coherent scattering has a strong energy dependence, it is thought that these delayed neutrinos will mostly appear at energies higher than the mass-delayed neutrinos, making this effect easier to identify.
To assess the effect of a neutrino echoes on the derived mass limit, we use the neutrino echo data from Gullin \textit{et al} \cite{Gullin2022-ez}, which uses the same solar-metallicity 40$M_{\odot}$ progenitor (s40) \cite{Woosley2007-nz} evolved with \texttt{GR1D} code \cite{OConnor2010-gv, OConnor2015-ju} instead of \texttt{PROMETHEUS-VERTEX} code. First, we introduce a neutrino mass delay into a signal with a neutrino echo, and then we compare this to a signal without a neutrino echo. Specifically, by carrying out a log-likelihood fit similar to the one described in Sec. IIIB, although a neutrino echo can 'fake' a larger neutrino mass delay, it has a noticeably larger absolute $\chi^2$, showing that this feature is distinguishable. Further studies are needed to derive a mass limit in the presence of a neutrino echo. We speculate that developing a more sophisticated fit which additional parameters to approximate the neutrino echo could break any degeneracy between these effects.\par
Based on this survey of possible effects superimposed on the ToF signature, we conclude that they are subdominant effects that can either be neglected or recognized and potentially included in the $m_\nu$ analysis. This corroborates that the characteristic energy-dependent delay beyond the sharp neutrino cutoff is a well-recognizable signature that can be exploited to set a tight neutrino mass limit with only a minimum of assumptions on the original neutrino signal required. 

\section{\label{sec:level12}Conclusion}

In this work, we have derived a mass sensitivity of $0.39^{+0.06}_{-0.01}$ \unit{\electronvolt} for a 95\% CL bound using neutrinos from a black hole-forming core-collapse supernova detected in a JUNO-like detector. This limit is tighter than the current constraint of 0.8 \unit{\electronvolt} (90\% CL) from KATRIN \cite{The_KATRIN_Collaboration2022-jx}. We have also shown that for a nearby or more luminous CCSN, this limit could be competitive with the target limit from KATRIN of 0.2 \unit{\electronvolt} (90\% CL) \cite{KATRIN:2005fny}, and with THEIA-100, this limit could be far exceeded. At the very least, a high-statistics observation of this event would allow a model-independent limit to be set on this value.\par
We have shown in this paper that the combination of a ToF analysis and liquid scintillator detectors has exceptional potential, but it is not unique. For instance, a similar analysis with CC argon reactions in DUNE \cite{DUNE_Collaboration2015-sb} could derive a limit on the $\Pnue$ mass using exactly the prescription detailed in this work. Complimentary analyses across many detectors and locations could tighten the $\nu$ mass limit even further. We are excited by the ever-growing supernova community \cite{Kharusi2020-bk, Al_Kharusi2021-uf}. With some luck, JUNO and other next-generation detectors will be running when neutrinos from the next core-collapse supernova reach Earth, and we can use these cosmic messengers to test the huge theoretical and experimental advances we have made since SN1987A. 

\begin{acknowledgments}
We thank David Maksimovic and Vsevolod Orekhov for discussions throughout the course of this work. We would also like to thank Hans-Thomas Janka and Daniel Kresse for their help in understanding CCSN mechanisms and simulations, Evan O'Connor for his help in understanding neutrino echoes, and Scott Oser for his help with statistics. We are grateful to Yufeng Li and Mariangela Settimo for carefully proofreading the manuscript. This work is supported by PRISMA+ Cluster of Excellence.
\end{acknowledgments}
\newpage
\appendix

\bibliographystyle{apsrev4-2}
\bibliography{ref.bib}

\begin{thebibliography}{80}%
\makeatletter
\providecommand \@ifxundefined [1]{%
 \@ifx{#1\undefined}
}%
\providecommand \@ifnum [1]{%
 \ifnum #1\expandafter \@firstoftwo
 \else \expandafter \@secondoftwo
 \fi
}%
\providecommand \@ifx [1]{%
 \ifx #1\expandafter \@firstoftwo
 \else \expandafter \@secondoftwo
 \fi
}%
\providecommand \natexlab [1]{#1}%
\providecommand \enquote  [1]{``#1''}%
\providecommand \bibnamefont  [1]{#1}%
\providecommand \bibfnamefont [1]{#1}%
\providecommand \citenamefont [1]{#1}%
\providecommand \href@noop [0]{\@secondoftwo}%
\providecommand \href [0]{\begingroup \@sanitize@url \@href}%
\providecommand \@href[1]{\@@startlink{#1}\@@href}%
\providecommand \@@href[1]{\endgroup#1\@@endlink}%
\providecommand \@sanitize@url [0]{\catcode `\\12\catcode `\$12\catcode
  `\&12\catcode `\#12\catcode `\^12\catcode `\_12\catcode `\%12\relax}%
\providecommand \@@startlink[1]{}%
\providecommand \@@endlink[0]{}%
\providecommand \url  [0]{\begingroup\@sanitize@url \@url }%
\providecommand \@url [1]{\endgroup\@href {#1}{\urlprefix }}%
\providecommand \urlprefix  [0]{URL }%
\providecommand \Eprint [0]{\href }%
\providecommand \doibase [0]{https://doi.org/}%
\providecommand \selectlanguage [0]{\@gobble}%
\providecommand \bibinfo  [0]{\@secondoftwo}%
\providecommand \bibfield  [0]{\@secondoftwo}%
\providecommand \translation [1]{[#1]}%
\providecommand \BibitemOpen [0]{}%
\providecommand \bibitemStop [0]{}%
\providecommand \bibitemNoStop [0]{.\EOS\space}%
\providecommand \EOS [0]{\spacefactor3000\relax}%
\providecommand \BibitemShut  [1]{\csname bibitem#1\endcsname}%
\let\auto@bib@innerbib\@empty
\bibitem [{\citenamefont {{Super Kamiokande Collaboration}}\ \emph
  {et~al.}(1998)\citenamefont {{Super Kamiokande Collaboration}}, \citenamefont
  {Fukuda}, \citenamefont {Hayakawa}, \citenamefont {Ichihara}, \citenamefont
  {Inoue},\ and\ \citenamefont {{\textit{et al}}}}]{Fukuda1998-jq}%
  \BibitemOpen
  \bibfield  {author} {\bibinfo {author} {\bibnamefont {{Super Kamiokande
  Collaboration}}}, \bibinfo {author} {\bibfnamefont {Y.}~\bibnamefont
  {Fukuda}}, \bibinfo {author} {\bibfnamefont {T.}~\bibnamefont {Hayakawa}},
  \bibinfo {author} {\bibfnamefont {E.}~\bibnamefont {Ichihara}}, \bibinfo
  {author} {\bibfnamefont {K.}~\bibnamefont {Inoue}},\ and\ \bibinfo {author}
  {\bibnamefont {{\textit{et al}}}},\ }\href@noop {} {\bibfield  {journal}
  {\bibinfo  {journal} {Phys. Rev. Lett.}\ }\textbf {\bibinfo {volume} {81}},\
  \bibinfo {pages} {1562} (\bibinfo {year} {1998})}\BibitemShut {NoStop}%
\bibitem [{\citenamefont {{SNO Collaboration}}\ \emph
  {et~al.}(2002)\citenamefont {{SNO Collaboration}}, \citenamefont {Ahmad},\
  and\ \citenamefont {{\textit{et al}}}}]{Ahmad2002-xy}%
  \BibitemOpen
  \bibfield  {author} {\bibinfo {author} {\bibnamefont {{SNO Collaboration}}},
  \bibinfo {author} {\bibfnamefont {Q.~R.}\ \bibnamefont {Ahmad}},\ and\
  \bibinfo {author} {\bibnamefont {{\textit{et al}}}},\ }\href@noop {}
  {\bibfield  {journal} {\bibinfo  {journal} {Phys. Rev. Lett.}\ }\textbf
  {\bibinfo {volume} {89}},\ \bibinfo {pages} {011301} (\bibinfo {year}
  {2002})}\BibitemShut {NoStop}%
\bibitem [{\citenamefont {Zatsepin}(1968)}]{Zatsepin:1968kt}%
  \BibitemOpen
  \bibfield  {author} {\bibinfo {author} {\bibfnamefont {G.~T.}\ \bibnamefont
  {Zatsepin}},\ }\href@noop {} {\bibfield  {journal} {\bibinfo  {journal}
  {Pisma Zh. Eksp. Teor. Fiz.}\ }\textbf {\bibinfo {volume} {8}},\ \bibinfo
  {pages} {333} (\bibinfo {year} {1968})}\BibitemShut {NoStop}%
\bibitem [{\citenamefont {Stodolsky}(2000)}]{Stodolsky2000-ab}%
  \BibitemOpen
  \bibfield  {author} {\bibinfo {author} {\bibfnamefont {L.}~\bibnamefont
  {Stodolsky}},\ }\href@noop {} {\bibfield  {journal} {\bibinfo  {journal}
  {Phys. Lett. B}\ }\textbf {\bibinfo {volume} {473}},\ \bibinfo {pages} {61}
  (\bibinfo {year} {2000})}\BibitemShut {NoStop}%
\bibitem [{Fan(2016)}]{Fanizza2016-id}%
  \BibitemOpen
  \href@noop {} {\bibfield  {journal} {\bibinfo  {journal} {Phys. Lett. B}\
  }\textbf {\bibinfo {volume} {757}},\ \bibinfo {pages} {505} (\bibinfo {year}
  {2016})}\BibitemShut {NoStop}%
\bibitem [{\citenamefont {Fleury}(2016)}]{Fleury2016-lj}%
  \BibitemOpen
  \bibfield  {author} {\bibinfo {author} {\bibfnamefont {P.}~\bibnamefont
  {Fleury}},\ }\href@noop {} {\bibfield  {journal} {\bibinfo  {journal} {Phys.
  Lett. B}\ }\textbf {\bibinfo {volume} {760}},\ \bibinfo {pages} {350}
  (\bibinfo {year} {2016})}\BibitemShut {NoStop}%
\bibitem [{\citenamefont {Beacom}\ \emph {et~al.}(2000)\citenamefont {Beacom},
  \citenamefont {Boyd},\ and\ \citenamefont {Mezzacappa}}]{Beacom2000-gp}%
  \BibitemOpen
  \bibfield  {author} {\bibinfo {author} {\bibfnamefont {J.~F.}\ \bibnamefont
  {Beacom}}, \bibinfo {author} {\bibfnamefont {R.~N.}\ \bibnamefont {Boyd}},\
  and\ \bibinfo {author} {\bibfnamefont {A.}~\bibnamefont {Mezzacappa}},\
  }\href@noop {} {\bibfield  {journal} {\bibinfo  {journal} {Phys. Rev. Lett.}\
  }\textbf {\bibinfo {volume} {85}},\ \bibinfo {pages} {3568} (\bibinfo {year}
  {2000})}\BibitemShut {NoStop}%
\bibitem [{\citenamefont {Beacom}\ \emph {et~al.}(2001)\citenamefont {Beacom},
  \citenamefont {Boyd},\ and\ \citenamefont {Mezzacappa}}]{Beacom2001-qf}%
  \BibitemOpen
  \bibfield  {author} {\bibinfo {author} {\bibfnamefont {J.~F.}\ \bibnamefont
  {Beacom}}, \bibinfo {author} {\bibfnamefont {R.~N.}\ \bibnamefont {Boyd}},\
  and\ \bibinfo {author} {\bibfnamefont {A.}~\bibnamefont {Mezzacappa}},\
  }\href@noop {} {\bibfield  {journal} {\bibinfo  {journal} {Phys. Rev. D Part.
  Fields}\ }\textbf {\bibinfo {volume} {63}} (\bibinfo {year}
  {2001})}\BibitemShut {NoStop}%
\bibitem [{\citenamefont {Hirata}\ \emph {et~al.}(1987)\citenamefont {Hirata},
  \citenamefont {Kajita}, \citenamefont {Koshiba}, \citenamefont {Nakahata},\
  and\ \citenamefont {Oyama}}]{Hirata1987-pw}%
  \BibitemOpen
  \bibfield  {author} {\bibinfo {author} {\bibfnamefont {K.}~\bibnamefont
  {Hirata}}, \bibinfo {author} {\bibfnamefont {T.}~\bibnamefont {Kajita}},
  \bibinfo {author} {\bibfnamefont {M.}~\bibnamefont {Koshiba}}, \bibinfo
  {author} {\bibfnamefont {M.}~\bibnamefont {Nakahata}},\ and\ \bibinfo
  {author} {\bibfnamefont {Y.~t.}\ \bibnamefont {Oyama}},\ }\href@noop {}
  {\bibfield  {journal} {\bibinfo  {journal} {Phys. Rev. Lett.}\ }\textbf
  {\bibinfo {volume} {58}},\ \bibinfo {pages} {1490} (\bibinfo {year}
  {1987})}\BibitemShut {NoStop}%
\bibitem [{Bio(1987)}]{Bionta1987-so}%
  \BibitemOpen
  \href@noop {} {\bibfield  {journal} {\bibinfo  {journal} {Phys. Rev. Lett.}\
  }\textbf {\bibinfo {volume} {58}},\ \bibinfo {pages} {1494} (\bibinfo {year}
  {1987})}\BibitemShut {NoStop}%
\bibitem [{\citenamefont {Alexeyev}\ \emph {et~al.}(1988)\citenamefont
  {Alexeyev}, \citenamefont {Alexeyeva}, \citenamefont {Krivosheina},\ and\
  \citenamefont {Volchenko}}]{Alexeyev1988-ts}%
  \BibitemOpen
  \bibfield  {author} {\bibinfo {author} {\bibfnamefont {E.~N.}\ \bibnamefont
  {Alexeyev}}, \bibinfo {author} {\bibfnamefont {L.~N.}\ \bibnamefont
  {Alexeyeva}}, \bibinfo {author} {\bibfnamefont {I.~V.}\ \bibnamefont
  {Krivosheina}},\ and\ \bibinfo {author} {\bibfnamefont {V.~I.}\ \bibnamefont
  {Volchenko}},\ }\href@noop {} {\bibfield  {journal} {\bibinfo  {journal}
  {Phys. Lett. B}\ }\textbf {\bibinfo {volume} {205}},\ \bibinfo {pages} {209}
  (\bibinfo {year} {1988})}\BibitemShut {NoStop}%
\bibitem [{\citenamefont {Arnett}\ and\ \citenamefont
  {Rosner}(1987)}]{Arnett1987-zg}%
  \BibitemOpen
  \bibfield  {author} {\bibinfo {author} {\bibfnamefont {W.~D.}\ \bibnamefont
  {Arnett}}\ and\ \bibinfo {author} {\bibfnamefont {J.~L.}\ \bibnamefont
  {Rosner}},\ }\href@noop {} {\bibfield  {journal} {\bibinfo  {journal} {Phys.
  Rev. Lett.}\ }\textbf {\bibinfo {volume} {58}},\ \bibinfo {pages} {1906}
  (\bibinfo {year} {1987})}\BibitemShut {NoStop}%
\bibitem [{\citenamefont {Loredo}\ and\ \citenamefont
  {Lamb}(2002)}]{Loredo2002-fr}%
  \BibitemOpen
  \bibfield  {author} {\bibinfo {author} {\bibfnamefont {T.~J.}\ \bibnamefont
  {Loredo}}\ and\ \bibinfo {author} {\bibfnamefont {D.~Q.}\ \bibnamefont
  {Lamb}},\ }\href@noop {} {\bibfield  {journal} {\bibinfo  {journal} {Phys.
  Rev. D Part. Fields}\ }\textbf {\bibinfo {volume} {65}} (\bibinfo {year}
  {2002})}\BibitemShut {NoStop}%
\bibitem [{\citenamefont {Pagliaroli}\ \emph {et~al.}(2010)\citenamefont
  {Pagliaroli}, \citenamefont {Rossi-Torres},\ and\ \citenamefont
  {Vissani}}]{Pagliaroli2010-do}%
  \BibitemOpen
  \bibfield  {author} {\bibinfo {author} {\bibfnamefont {G.}~\bibnamefont
  {Pagliaroli}}, \bibinfo {author} {\bibfnamefont {F.}~\bibnamefont
  {Rossi-Torres}},\ and\ \bibinfo {author} {\bibfnamefont {F.}~\bibnamefont
  {Vissani}},\ }\href@noop {} {\bibfield  {journal} {\bibinfo  {journal}
  {Astropart. Phys.}\ }\textbf {\bibinfo {volume} {33}},\ \bibinfo {pages}
  {287} (\bibinfo {year} {2010})}\BibitemShut {NoStop}%
\bibitem [{\citenamefont {Abe}\ \emph {et~al.}(2016)\citenamefont {Abe},
  \citenamefont {Haga}, \citenamefont {Hayato}, \citenamefont {Ikeda},\ and\
  \citenamefont {Iyogi}}]{Abe2016-zo}%
  \BibitemOpen
  \bibfield  {author} {\bibinfo {author} {\bibfnamefont {K.}~\bibnamefont
  {Abe}}, \bibinfo {author} {\bibfnamefont {Y.}~\bibnamefont {Haga}}, \bibinfo
  {author} {\bibfnamefont {Y.}~\bibnamefont {Hayato}}, \bibinfo {author}
  {\bibfnamefont {M.}~\bibnamefont {Ikeda}},\ and\ \bibinfo {author}
  {\bibfnamefont {K.~t.~S.}\ \bibnamefont {Iyogi}},\ }\href@noop {} {\bibfield
  {journal} {\bibinfo  {journal} {Phys. Rev. D.}\ }\textbf {\bibinfo {volume}
  {94}} (\bibinfo {year} {2016})}\BibitemShut {NoStop}%
\bibitem [{\citenamefont {Lu}\ \emph {et~al.}(2015)\citenamefont {Lu},
  \citenamefont {Cao}, \citenamefont {Li},\ and\ \citenamefont
  {Zhou}}]{Lu2015-uu}%
  \BibitemOpen
  \bibfield  {author} {\bibinfo {author} {\bibfnamefont {J.-S.}\ \bibnamefont
  {Lu}}, \bibinfo {author} {\bibfnamefont {J.}~\bibnamefont {Cao}}, \bibinfo
  {author} {\bibfnamefont {Y.-F.}\ \bibnamefont {Li}},\ and\ \bibinfo {author}
  {\bibfnamefont {S.}~\bibnamefont {Zhou}},\ }\href@noop {} {\bibfield
  {journal} {\bibinfo  {journal} {J. Cosmol. Astropart. Phys.}\ }\textbf
  {\bibinfo {volume} {2015}}\bibinfo  {number} { (05)},\ \bibinfo {pages}
  {044}}\BibitemShut {NoStop}%
\bibitem [{\citenamefont {An}(2016)}]{An2016-kb}%
  \BibitemOpen
\bibfield  {number} {  }\bibfield  {author} {\bibinfo {author} {\bibfnamefont
  {F.}~\bibnamefont {An}},\ }\href@noop {} {\bibfield  {journal} {\bibinfo
  {journal} {J. Phys. G Nucl. Part. Phys.}\ }\textbf {\bibinfo {volume} {43}},\
  \bibinfo {pages} {030401} (\bibinfo {year} {2016})}\BibitemShut {NoStop}%
\bibitem [{\citenamefont {Scholberg}(2018)}]{Scholberg2018-ms}%
  \BibitemOpen
  \bibfield  {author} {\bibinfo {author} {\bibfnamefont {K.}~\bibnamefont
  {Scholberg}},\ }\href@noop {} {\bibfield  {journal} {\bibinfo  {journal} {J.
  Phys. G Nucl. Part. Phys.}\ }\textbf {\bibinfo {volume} {45}},\ \bibinfo
  {pages} {014002} (\bibinfo {year} {2018})}\BibitemShut {NoStop}%
\bibitem [{\citenamefont {Choudhury}\ and\ \citenamefont
  {Choubey}(2018)}]{Choudhury2018-jh}%
  \BibitemOpen
  \bibfield  {author} {\bibinfo {author} {\bibfnamefont {S.~R.}\ \bibnamefont
  {Choudhury}}\ and\ \bibinfo {author} {\bibfnamefont {S.}~\bibnamefont
  {Choubey}},\ }\href@noop {} {\bibfield  {journal} {\bibinfo  {journal} {J.
  Cosmol. Astropart. Phys.}\ }\textbf {\bibinfo {volume} {2018}}\bibinfo
  {number} { (09)},\ \bibinfo {pages} {017}}\BibitemShut {NoStop}%
\bibitem [{\citenamefont {{Planck Collaboration}}\ \emph
  {et~al.}(2020)\citenamefont {{Planck Collaboration}}, \citenamefont
  {Aghanim}, \citenamefont {Akrami}, \citenamefont {Ashdown}, \citenamefont
  {Aumont},\ and\ \citenamefont {Baccigalupi}}]{Planck_Collaboration2020-vu}%
  \BibitemOpen
\bibfield  {number} {  }\bibfield  {author} {\bibinfo {author} {\bibnamefont
  {{Planck Collaboration}}}, \bibinfo {author} {\bibfnamefont {N.}~\bibnamefont
  {Aghanim}}, \bibinfo {author} {\bibfnamefont {Y.}~\bibnamefont {Akrami}},
  \bibinfo {author} {\bibfnamefont {M.}~\bibnamefont {Ashdown}}, \bibinfo
  {author} {\bibfnamefont {J.}~\bibnamefont {Aumont}},\ and\ \bibinfo {author}
  {\bibfnamefont {t.}~\bibnamefont {Baccigalupi}, \bibfnamefont {C}},\
  }\href@noop {} {\bibfield  {journal} {\bibinfo  {journal} {Astron.
  Astrophys.}\ }\textbf {\bibinfo {volume} {641}},\ \bibinfo {pages} {A6}
  (\bibinfo {year} {2020})}\BibitemShut {NoStop}%
\bibitem [{\citenamefont {Di~Valentino}\ \emph {et~al.}(2015)\citenamefont
  {Di~Valentino}, \citenamefont {Melchiorri},\ and\ \citenamefont
  {Silk}}]{Di_Valentino2015-he}%
  \BibitemOpen
  \bibfield  {author} {\bibinfo {author} {\bibfnamefont {E.}~\bibnamefont
  {Di~Valentino}}, \bibinfo {author} {\bibfnamefont {A.}~\bibnamefont
  {Melchiorri}},\ and\ \bibinfo {author} {\bibfnamefont {J.}~\bibnamefont
  {Silk}},\ }\href@noop {} {\bibfield  {journal} {\bibinfo  {journal} {Phys.
  Rev. D}\ }\textbf {\bibinfo {volume} {92}} (\bibinfo {year}
  {2015})}\BibitemShut {NoStop}%
\bibitem [{\citenamefont {{The KATRIN Collaboration}}\ \emph
  {et~al.}(2022)\citenamefont {{The KATRIN Collaboration}}, \citenamefont
  {Aker}, \citenamefont {Beglarian}, \citenamefont {Behrens}, \citenamefont
  {Berlev},\ and\ \citenamefont {Besserer}}]{The_KATRIN_Collaboration2022-jx}%
  \BibitemOpen
  \bibfield  {author} {\bibinfo {author} {\bibnamefont {{The KATRIN
  Collaboration}}}, \bibinfo {author} {\bibfnamefont {M.}~\bibnamefont {Aker}},
  \bibinfo {author} {\bibfnamefont {A.}~\bibnamefont {Beglarian}}, \bibinfo
  {author} {\bibfnamefont {J.}~\bibnamefont {Behrens}}, \bibinfo {author}
  {\bibfnamefont {A.}~\bibnamefont {Berlev}},\ and\ \bibinfo {author}
  {\bibfnamefont {t.}~\bibnamefont {Besserer}, \bibfnamefont {U}},\ }\href@noop
  {} {\bibfield  {journal} {\bibinfo  {journal} {Nat. Phys.}\ }\textbf
  {\bibinfo {volume} {18}},\ \bibinfo {pages} {160} (\bibinfo {year}
  {2022})}\BibitemShut {NoStop}%
\bibitem [{\citenamefont {Angrik}\ \emph {et~al.}(2005)\citenamefont {Angrik}
  \emph {et~al.}}]{KATRIN:2005fny}%
  \BibitemOpen
  \bibfield  {author} {\bibinfo {author} {\bibfnamefont {J.}~\bibnamefont
  {Angrik}} \emph {et~al.} (\bibinfo {collaboration} {KATRIN}),\ }\href@noop {}
  {\  (\bibinfo {year} {2005})}\BibitemShut {NoStop}%
\bibitem [{\citenamefont {Esfahani}\ \emph {et~al.}(2017)\citenamefont
  {Esfahani}, \citenamefont {Asner}, \citenamefont {B{\"o}ser},\ and\
  \citenamefont {Cervantes}}]{Esfahani2017-og}%
  \BibitemOpen
  \bibfield  {author} {\bibinfo {author} {\bibfnamefont {A.~A.}\ \bibnamefont
  {Esfahani}}, \bibinfo {author} {\bibfnamefont {D.~M.}\ \bibnamefont {Asner}},
  \bibinfo {author} {\bibfnamefont {S.}~\bibnamefont {B{\"o}ser}},\ and\
  \bibinfo {author} {\bibfnamefont {R.~t.}\ \bibnamefont {Cervantes}},\
  }\href@noop {} {\bibfield  {journal} {\bibinfo  {journal} {J. Phys. G Nucl.
  Part. Phys.}\ }\textbf {\bibinfo {volume} {44}},\ \bibinfo {pages} {054004}
  (\bibinfo {year} {2017})}\BibitemShut {NoStop}%
\bibitem [{\citenamefont {Adams}\ \emph
  {et~al.}(2013{\natexlab{a}})\citenamefont {Adams}, \citenamefont {Kochanek},
  \citenamefont {Beacom}, \citenamefont {Vagins},\ and\ \citenamefont
  {Stanek}}]{Adams2013-yj}%
  \BibitemOpen
  \bibfield  {author} {\bibinfo {author} {\bibfnamefont {S.~M.}\ \bibnamefont
  {Adams}}, \bibinfo {author} {\bibfnamefont {C.~S.}\ \bibnamefont {Kochanek}},
  \bibinfo {author} {\bibfnamefont {J.~F.}\ \bibnamefont {Beacom}}, \bibinfo
  {author} {\bibfnamefont {M.~R.}\ \bibnamefont {Vagins}},\ and\ \bibinfo
  {author} {\bibfnamefont {K.~Z.}\ \bibnamefont {Stanek}},\ }\href@noop {}
  {\bibfield  {journal} {\bibinfo  {journal} {Astrophys. J.}\ }\textbf
  {\bibinfo {volume} {778}},\ \bibinfo {pages} {164} (\bibinfo {year}
  {2013}{\natexlab{a}})}\BibitemShut {NoStop}%
\bibitem [{\citenamefont {Rozwadowska}\ \emph {et~al.}(2021)\citenamefont
  {Rozwadowska}, \citenamefont {Vissani},\ and\ \citenamefont
  {Cappellaro}}]{Rozwadowska2021-vc}%
  \BibitemOpen
  \bibfield  {author} {\bibinfo {author} {\bibfnamefont {K.}~\bibnamefont
  {Rozwadowska}}, \bibinfo {author} {\bibfnamefont {F.}~\bibnamefont
  {Vissani}},\ and\ \bibinfo {author} {\bibfnamefont {E.}~\bibnamefont
  {Cappellaro}},\ }\href@noop {} {\bibfield  {journal} {\bibinfo  {journal}
  {New astron.}\ }\textbf {\bibinfo {volume} {83}},\ \bibinfo {pages} {101498}
  (\bibinfo {year} {2021})}\BibitemShut {NoStop}%
\bibitem [{\citenamefont {O'Connor}\ and\ \citenamefont
  {Ott}(2011)}]{OConnor2011-vl}%
  \BibitemOpen
  \bibfield  {author} {\bibinfo {author} {\bibfnamefont {E.}~\bibnamefont
  {O'Connor}}\ and\ \bibinfo {author} {\bibfnamefont {C.~D.}\ \bibnamefont
  {Ott}},\ }\href@noop {} {\bibfield  {journal} {\bibinfo  {journal}
  {Astrophys. J.}\ }\textbf {\bibinfo {volume} {730}},\ \bibinfo {pages} {70}
  (\bibinfo {year} {2011})}\BibitemShut {NoStop}%
\bibitem [{\citenamefont {Neustadt}\ \emph {et~al.}(2021)\citenamefont
  {Neustadt}, \citenamefont {Kochanek}, \citenamefont {Stanek}, \citenamefont
  {Basinger}, \citenamefont {Jayasinghe}, \citenamefont {Garling},
  \citenamefont {Adams},\ and\ \citenamefont {Gerke}}]{Neustadt2021-zj}%
  \BibitemOpen
  \bibfield  {author} {\bibinfo {author} {\bibfnamefont {J.~M.~M.}\
  \bibnamefont {Neustadt}}, \bibinfo {author} {\bibfnamefont {C.~S.}\
  \bibnamefont {Kochanek}}, \bibinfo {author} {\bibfnamefont {K.~Z.}\
  \bibnamefont {Stanek}}, \bibinfo {author} {\bibfnamefont {C.}~\bibnamefont
  {Basinger}}, \bibinfo {author} {\bibfnamefont {T.}~\bibnamefont
  {Jayasinghe}}, \bibinfo {author} {\bibfnamefont {C.~T.}\ \bibnamefont
  {Garling}}, \bibinfo {author} {\bibfnamefont {S.~M.}\ \bibnamefont {Adams}},\
  and\ \bibinfo {author} {\bibfnamefont {J.}~\bibnamefont {Gerke}},\
  }\href@noop {} {\bibfield  {journal} {\bibinfo  {journal} {Mon. Not. R.
  Astron. Soc.}\ }\textbf {\bibinfo {volume} {508}},\ \bibinfo {pages} {516}
  (\bibinfo {year} {2021})}\BibitemShut {NoStop}%
\bibitem [{\citenamefont {Janka}\ \emph {et~al.}(2012)\citenamefont {Janka},
  \citenamefont {Hanke}, \citenamefont {Huedepohl}, \citenamefont {Marek},
  \citenamefont {Mueller},\ and\ \citenamefont {Obergaulinger}}]{Janka2012-ds}%
  \BibitemOpen
  \bibfield  {author} {\bibinfo {author} {\bibfnamefont {H.-T.}\ \bibnamefont
  {Janka}}, \bibinfo {author} {\bibfnamefont {F.}~\bibnamefont {Hanke}},
  \bibinfo {author} {\bibfnamefont {L.}~\bibnamefont {Huedepohl}}, \bibinfo
  {author} {\bibfnamefont {A.}~\bibnamefont {Marek}}, \bibinfo {author}
  {\bibfnamefont {B.}~\bibnamefont {Mueller}},\ and\ \bibinfo {author}
  {\bibfnamefont {M.}~\bibnamefont {Obergaulinger}},\ }\href@noop {} {\
  (\bibinfo {year} {2012})}\BibitemShut {NoStop}%
\bibitem [{\citenamefont {Burrows}(2013)}]{Burrows2013-ca}%
  \BibitemOpen
  \bibfield  {author} {\bibinfo {author} {\bibfnamefont {A.}~\bibnamefont
  {Burrows}},\ }\href@noop {} {\bibfield  {journal} {\bibinfo  {journal} {Rev.
  Mod. Phys.}\ }\textbf {\bibinfo {volume} {85}},\ \bibinfo {pages} {245}
  (\bibinfo {year} {2013})}\BibitemShut {NoStop}%
\bibitem [{\citenamefont {Janka}(2017)}]{Janka2017-wn}%
  \BibitemOpen
  \bibfield  {author} {\bibinfo {author} {\bibfnamefont {H.-T.}\ \bibnamefont
  {Janka}},\ }in\ \href@noop {} {\emph {\bibinfo {booktitle} {Handbook of
  Supernovae}}}\ (\bibinfo  {publisher} {Springer International Publishing},\
  \bibinfo {address} {Cham},\ \bibinfo {year} {2017})\ pp.\ \bibinfo {pages}
  {1095--1150}\BibitemShut {NoStop}%
\bibitem [{\citenamefont {Li}\ \emph {et~al.}(2021)\citenamefont {Li},
  \citenamefont {Roberts},\ and\ \citenamefont {Beacom}}]{Li2021-no}%
  \BibitemOpen
  \bibfield  {author} {\bibinfo {author} {\bibfnamefont {S.~W.}\ \bibnamefont
  {Li}}, \bibinfo {author} {\bibfnamefont {L.~F.}\ \bibnamefont {Roberts}},\
  and\ \bibinfo {author} {\bibfnamefont {J.~F.}\ \bibnamefont {Beacom}},\
  }\href@noop {} {\bibfield  {journal} {\bibinfo  {journal} {Phys. Rev. D.}\
  }\textbf {\bibinfo {volume} {103}} (\bibinfo {year} {2021})}\BibitemShut
  {NoStop}%
\bibitem [{\citenamefont {Walk}\ \emph {et~al.}(2020)\citenamefont {Walk},
  \citenamefont {Tamborra}, \citenamefont {Janka}, \citenamefont {Summa},\ and\
  \citenamefont {Kresse}}]{Walk2020-pj}%
  \BibitemOpen
  \bibfield  {author} {\bibinfo {author} {\bibfnamefont {L.}~\bibnamefont
  {Walk}}, \bibinfo {author} {\bibfnamefont {I.}~\bibnamefont {Tamborra}},
  \bibinfo {author} {\bibfnamefont {H.-T.}\ \bibnamefont {Janka}}, \bibinfo
  {author} {\bibfnamefont {A.}~\bibnamefont {Summa}},\ and\ \bibinfo {author}
  {\bibfnamefont {D.}~\bibnamefont {Kresse}},\ }\href@noop {} {\bibfield
  {journal} {\bibinfo  {journal} {Phys. Rev. D.}\ }\textbf {\bibinfo {volume}
  {101}} (\bibinfo {year} {2020})}\BibitemShut {NoStop}%
\bibitem [{\citenamefont {Blondin}\ \emph {et~al.}(2003)\citenamefont
  {Blondin}, \citenamefont {Mezzacappa},\ and\ \citenamefont
  {DeMarino}}]{Blondin2003-sq}%
  \BibitemOpen
  \bibfield  {author} {\bibinfo {author} {\bibfnamefont {J.~M.}\ \bibnamefont
  {Blondin}}, \bibinfo {author} {\bibfnamefont {A.}~\bibnamefont
  {Mezzacappa}},\ and\ \bibinfo {author} {\bibfnamefont {C.}~\bibnamefont
  {DeMarino}},\ }\href@noop {} {\bibfield  {journal} {\bibinfo  {journal}
  {Astrophys. J.}\ }\textbf {\bibinfo {volume} {584}},\ \bibinfo {pages} {971}
  (\bibinfo {year} {2003})}\BibitemShut {NoStop}%
\bibitem [{Gar()}]{Garching}%
  \BibitemOpen
  \href@noop {} {\bibinfo {title} {{The Garching Core-Collapse Supernova
  Archive}}},\ \bibinfo {howpublished}
  {\url{https://wwwmpa.mpa-garching.mpg.de/ccsnarchive/}}\BibitemShut {NoStop}%
\bibitem [{\citenamefont {{Super Kamiokande Collaboration}}\ and\ \citenamefont
  {Ikeda}(2021)}]{Ikeda2021-ar}%
  \BibitemOpen
  \bibfield  {author} {\bibinfo {author} {\bibnamefont {{Super Kamiokande
  Collaboration}}}\ and\ \bibinfo {author} {\bibfnamefont {M.}~\bibnamefont
  {Ikeda}},\ }\href@noop {} {\bibfield  {journal} {\bibinfo  {journal} {J.
  Phys. Conf. Ser.}\ }\textbf {\bibinfo {volume} {2156}},\ \bibinfo {pages}
  {012150} (\bibinfo {year} {2021})}\BibitemShut {NoStop}%
\bibitem [{\citenamefont {{Proto-Collaboration, Hyper-Kamiokande}}\ \emph
  {et~al.}(2018)\citenamefont {{Proto-Collaboration, Hyper-Kamiokande}}, ,\
  and\ \citenamefont {Abe}}]{Proto-Collaboration_Hyper-Kamiokande2018-zc}%
  \BibitemOpen
  \bibfield  {author} {\bibinfo {author} {\bibnamefont {{Proto-Collaboration,
  Hyper-Kamiokande}}}, ,\ and\ \bibinfo {author} {\bibfnamefont {K.~a.}\
  \bibnamefont {Abe}},\ }\href@noop {} {\  (\bibinfo {year}
  {2018})}\BibitemShut {NoStop}%
\bibitem [{JUN(2022)}]{JUNO2022}%
  \BibitemOpen
  \href {https://doi.org/10.1016/j.ppnp.2021.103927} {\bibfield  {journal}
  {\bibinfo  {journal} {Progress in Particle and Nuclear Physics}\ }\textbf
  {\bibinfo {volume} {123}},\ \bibinfo {pages} {103927} (\bibinfo {year}
  {2022})}\BibitemShut {NoStop}%
\bibitem [{\citenamefont {Askins}\ \emph {et~al.}(2020)\citenamefont {Askins},
  \citenamefont {Bagdasarian}, \citenamefont {Barros}, \citenamefont {Beier},\
  and\ \citenamefont {\textit{et al}}}]{Askins2020}%
  \BibitemOpen
  \bibfield  {author} {\bibinfo {author} {\bibfnamefont {M.}~\bibnamefont
  {Askins}}, \bibinfo {author} {\bibfnamefont {Z.}~\bibnamefont {Bagdasarian}},
  \bibinfo {author} {\bibfnamefont {N.}~\bibnamefont {Barros}}, \bibinfo
  {author} {\bibfnamefont {E.~W.}\ \bibnamefont {Beier}},\ and\ \bibinfo
  {author} {\bibfnamefont {E.~B.}\ \bibnamefont {\textit{et al}}},\ }\bibfield
  {journal} {\bibinfo  {journal} {The European Physical Journal C}\ }\textbf
  {\bibinfo {volume} {80}},\ \href
  {https://doi.org/10.1140/epjc/s10052-020-7977-8}
  {10.1140/epjc/s10052-020-7977-8} (\bibinfo {year} {2020})\BibitemShut
  {NoStop}%
\bibitem [{\citenamefont {Zsoldos}\ \emph {et~al.}(2022)\citenamefont
  {Zsoldos}, \citenamefont {Bagdasarian}, \citenamefont {Gann~Orebi},
  \citenamefont {Barna},\ and\ \citenamefont {Dye}}]{Zsoldos}%
  \BibitemOpen
  \bibfield  {author} {\bibinfo {author} {\bibfnamefont {S.}~\bibnamefont
  {Zsoldos}}, \bibinfo {author} {\bibfnamefont {Z.}~\bibnamefont
  {Bagdasarian}}, \bibinfo {author} {\bibfnamefont {G.~D.}\ \bibnamefont
  {Gann~Orebi}}, \bibinfo {author} {\bibfnamefont {A.}~\bibnamefont {Barna}},\
  and\ \bibinfo {author} {\bibfnamefont {S.}~\bibnamefont {Dye}},\ }\href
  {https://doi.org/10.1140/epjc/s10052-022-11106-1} {\bibfield  {journal}
  {\bibinfo  {journal} {Eur. Phys. J. C}\ }\textbf {\bibinfo {volume} {82}},\
  \bibinfo {pages} {1151} (\bibinfo {year} {2022})},\ \Eprint
  {https://arxiv.org/abs/2204.12278} {arXiv:2204.12278 [hep-ex]} \BibitemShut
  {NoStop}%
\bibitem [{\citenamefont {Woosley}\ and\ \citenamefont
  {Heger}(2007)}]{Woosley2007-nz}%
  \BibitemOpen
  \bibfield  {author} {\bibinfo {author} {\bibfnamefont {S.}~\bibnamefont
  {Woosley}}\ and\ \bibinfo {author} {\bibfnamefont {A.}~\bibnamefont
  {Heger}},\ }\href@noop {} {\bibfield  {journal} {\bibinfo  {journal} {Phys.
  Rep.}\ }\textbf {\bibinfo {volume} {442}},\ \bibinfo {pages} {269} (\bibinfo
  {year} {2007})}\BibitemShut {NoStop}%
\bibitem [{\citenamefont {Woosley}\ \emph {et~al.}(2002)\citenamefont
  {Woosley}, \citenamefont {Heger},\ and\ \citenamefont
  {Weaver}}]{Woosley2002-xx}%
  \BibitemOpen
  \bibfield  {author} {\bibinfo {author} {\bibfnamefont {S.~E.}\ \bibnamefont
  {Woosley}}, \bibinfo {author} {\bibfnamefont {A.}~\bibnamefont {Heger}},\
  and\ \bibinfo {author} {\bibfnamefont {T.~A.}\ \bibnamefont {Weaver}},\
  }\href@noop {} {\bibfield  {journal} {\bibinfo  {journal} {Rev. Mod. Phys.}\
  }\textbf {\bibinfo {volume} {74}},\ \bibinfo {pages} {1015} (\bibinfo {year}
  {2002})}\BibitemShut {NoStop}%
\bibitem [{\citenamefont {Rampp}\ and\ \citenamefont
  {Janka}(2002)}]{Rampp2002-qw}%
  \BibitemOpen
  \bibfield  {author} {\bibinfo {author} {\bibfnamefont {M.}~\bibnamefont
  {Rampp}}\ and\ \bibinfo {author} {\bibfnamefont {H.-T.}\ \bibnamefont
  {Janka}},\ }\href@noop {} {\  (\bibinfo {year} {2002})}\BibitemShut {NoStop}%
\bibitem [{\citenamefont {Marek}\ \emph {et~al.}(2005)\citenamefont {Marek},
  \citenamefont {Dimmelmeier}, \citenamefont {Janka}, \citenamefont {Mueller},\
  and\ \citenamefont {Buras}}]{Marek2005-if}%
  \BibitemOpen
  \bibfield  {author} {\bibinfo {author} {\bibfnamefont {A.}~\bibnamefont
  {Marek}}, \bibinfo {author} {\bibfnamefont {H.}~\bibnamefont {Dimmelmeier}},
  \bibinfo {author} {\bibfnamefont {H.-T.}\ \bibnamefont {Janka}}, \bibinfo
  {author} {\bibfnamefont {E.}~\bibnamefont {Mueller}},\ and\ \bibinfo {author}
  {\bibfnamefont {R.}~\bibnamefont {Buras}},\ }\href@noop {} {\  (\bibinfo
  {year} {2005})}\BibitemShut {NoStop}%
\bibitem [{\citenamefont {Keil}\ \emph {et~al.}(2003)\citenamefont {Keil},
  \citenamefont {Raffelt},\ and\ \citenamefont {Janka}}]{Keil2003-kf}%
  \BibitemOpen
  \bibfield  {author} {\bibinfo {author} {\bibfnamefont {M.~T.}\ \bibnamefont
  {Keil}}, \bibinfo {author} {\bibfnamefont {G.~G.}\ \bibnamefont {Raffelt}},\
  and\ \bibinfo {author} {\bibfnamefont {H.-T.}\ \bibnamefont {Janka}},\
  }\href@noop {} {\bibfield  {journal} {\bibinfo  {journal} {Astrophys. J.}\
  }\textbf {\bibinfo {volume} {590}},\ \bibinfo {pages} {971} (\bibinfo {year}
  {2003})}\BibitemShut {NoStop}%
\bibitem [{\citenamefont {Tamborra}\ \emph {et~al.}(2012)\citenamefont
  {Tamborra}, \citenamefont {M{\"u}ller}, \citenamefont {H{\"u}depohl},
  \citenamefont {Janka},\ and\ \citenamefont {Raffelt}}]{Tamborra2012-vu}%
  \BibitemOpen
  \bibfield  {author} {\bibinfo {author} {\bibfnamefont {I.}~\bibnamefont
  {Tamborra}}, \bibinfo {author} {\bibfnamefont {B.}~\bibnamefont
  {M{\"u}ller}}, \bibinfo {author} {\bibfnamefont {L.}~\bibnamefont
  {H{\"u}depohl}}, \bibinfo {author} {\bibfnamefont {H.-T.}\ \bibnamefont
  {Janka}},\ and\ \bibinfo {author} {\bibfnamefont {G.}~\bibnamefont
  {Raffelt}},\ }\href@noop {} {\bibfield  {journal} {\bibinfo  {journal} {Phys.
  rev.}\ }\textbf {\bibinfo {volume} {86}} (\bibinfo {year}
  {2012})}\BibitemShut {NoStop}%
\bibitem [{\citenamefont {Wolfenstein}(1978)}]{Wolfenstein1978-wy}%
  \BibitemOpen
  \bibfield  {author} {\bibinfo {author} {\bibfnamefont {L.}~\bibnamefont
  {Wolfenstein}},\ }\href@noop {} {\bibfield  {journal} {\bibinfo  {journal}
  {Phys. Rev. D Part. Fields}\ }\textbf {\bibinfo {volume} {17}},\ \bibinfo
  {pages} {2369} (\bibinfo {year} {1978})}\BibitemShut {NoStop}%
\bibitem [{\citenamefont {Wolfenstein}(1979)}]{Wolfenstein1979-cc}%
  \BibitemOpen
  \bibfield  {author} {\bibinfo {author} {\bibfnamefont {L.}~\bibnamefont
  {Wolfenstein}},\ }\href@noop {} {\bibfield  {journal} {\bibinfo  {journal}
  {Phys. Rev. D Part. Fields}\ }\textbf {\bibinfo {volume} {20}},\ \bibinfo
  {pages} {2634} (\bibinfo {year} {1979})}\BibitemShut {NoStop}%
\bibitem [{\citenamefont {Dighe}\ and\ \citenamefont
  {Smirnov}(2000)}]{Dighe2000-rv}%
  \BibitemOpen
  \bibfield  {author} {\bibinfo {author} {\bibfnamefont {A.~S.}\ \bibnamefont
  {Dighe}}\ and\ \bibinfo {author} {\bibfnamefont {A.~Y.}\ \bibnamefont
  {Smirnov}},\ }\href@noop {} {\bibfield  {journal} {\bibinfo  {journal} {Phys.
  Rev. D Part. Fields}\ }\textbf {\bibinfo {volume} {62}} (\bibinfo {year}
  {2000})}\BibitemShut {NoStop}%
\bibitem [{\citenamefont {Takahashi}\ \emph {et~al.}(2001)\citenamefont
  {Takahashi}, \citenamefont {Watanabe}, \citenamefont {Sato},\ and\
  \citenamefont {Totani}}]{Takahashi2001-ah}%
  \BibitemOpen
  \bibfield  {author} {\bibinfo {author} {\bibfnamefont {K.}~\bibnamefont
  {Takahashi}}, \bibinfo {author} {\bibfnamefont {M.}~\bibnamefont {Watanabe}},
  \bibinfo {author} {\bibfnamefont {K.}~\bibnamefont {Sato}},\ and\ \bibinfo
  {author} {\bibfnamefont {T.}~\bibnamefont {Totani}},\ }\href@noop {}
  {\bibfield  {journal} {\bibinfo  {journal} {Phys. Rev. D Part. Fields}\
  }\textbf {\bibinfo {volume} {64}} (\bibinfo {year} {2001})}\BibitemShut
  {NoStop}%
\bibitem [{\citenamefont {Kawagoe}\ \emph {et~al.}(2010)\citenamefont
  {Kawagoe}, \citenamefont {Yoshida}, \citenamefont {Kajino}, \citenamefont
  {Suzuki}, \citenamefont {Sumiyoshi},\ and\ \citenamefont
  {Yamada}}]{Kawagoe2010-hh}%
  \BibitemOpen
  \bibfield  {author} {\bibinfo {author} {\bibfnamefont {S.}~\bibnamefont
  {Kawagoe}}, \bibinfo {author} {\bibfnamefont {T.}~\bibnamefont {Yoshida}},
  \bibinfo {author} {\bibfnamefont {T.}~\bibnamefont {Kajino}}, \bibinfo
  {author} {\bibfnamefont {H.}~\bibnamefont {Suzuki}}, \bibinfo {author}
  {\bibfnamefont {K.}~\bibnamefont {Sumiyoshi}},\ and\ \bibinfo {author}
  {\bibfnamefont {S.}~\bibnamefont {Yamada}},\ }\href@noop {} {\bibfield
  {journal} {\bibinfo  {journal} {Phys. rev.}\ }\textbf {\bibinfo {volume}
  {81}} (\bibinfo {year} {2010})}\BibitemShut {NoStop}%
\bibitem [{\citenamefont {Lu}\ \emph {et~al.}(2016)\citenamefont {Lu},
  \citenamefont {Li},\ and\ \citenamefont {Zhou}}]{Lu2016-cv}%
  \BibitemOpen
  \bibfield  {author} {\bibinfo {author} {\bibfnamefont {J.-S.}\ \bibnamefont
  {Lu}}, \bibinfo {author} {\bibfnamefont {Y.-F.}\ \bibnamefont {Li}},\ and\
  \bibinfo {author} {\bibfnamefont {S.}~\bibnamefont {Zhou}},\ }\href@noop {}
  {\bibfield  {journal} {\bibinfo  {journal} {Phys. Rev. D.}\ }\textbf
  {\bibinfo {volume} {94}} (\bibinfo {year} {2016})}\BibitemShut {NoStop}%
\bibitem [{\citenamefont {Esteban}\ \emph {et~al.}(2020)\citenamefont
  {Esteban}, \citenamefont {Gonzalez-Garcia}, \citenamefont {Maltoni},
  \citenamefont {Schwetz},\ and\ \citenamefont {Zhou}}]{Esteban2020-ha}%
  \BibitemOpen
  \bibfield  {author} {\bibinfo {author} {\bibfnamefont {I.}~\bibnamefont
  {Esteban}}, \bibinfo {author} {\bibfnamefont {M.~C.}\ \bibnamefont
  {Gonzalez-Garcia}}, \bibinfo {author} {\bibfnamefont {M.}~\bibnamefont
  {Maltoni}}, \bibinfo {author} {\bibfnamefont {T.}~\bibnamefont {Schwetz}},\
  and\ \bibinfo {author} {\bibfnamefont {A.}~\bibnamefont {Zhou}},\ }\href@noop
  {} {\bibfield  {journal} {\bibinfo  {journal} {J. High Energy Phys.}\
  }\textbf {\bibinfo {volume} {2020}}\bibinfo  {number} { (9)}}\BibitemShut
  {NoStop}%
\bibitem [{\citenamefont {Strumia}\ and\ \citenamefont
  {Vissani}(2003)}]{Strumia2003-dx}%
  \BibitemOpen
\bibfield  {number} {  }\bibfield  {author} {\bibinfo {author} {\bibfnamefont
  {A.}~\bibnamefont {Strumia}}\ and\ \bibinfo {author} {\bibfnamefont
  {F.}~\bibnamefont {Vissani}},\ }\href@noop {} {\bibfield  {journal} {\bibinfo
   {journal} {Phys. Lett. B}\ }\textbf {\bibinfo {volume} {564}},\ \bibinfo
  {pages} {42} (\bibinfo {year} {2003})}\BibitemShut {NoStop}%
\bibitem [{\citenamefont {Scholberg}()}]{Scholberg-2021}%
  \BibitemOpen
  \bibfield  {author} {\bibinfo {author} {\bibfnamefont {K.}~\bibnamefont
  {Scholberg}},\ }\href@noop {} {\bibinfo {title} {Snowglobes: Supernova
  observatories with globes}},\ \bibinfo {howpublished} {GitHub, \url{
  https://github.com/SNOwGLoBES/snowglobes}}\BibitemShut {NoStop}%
\bibitem [{\citenamefont {Pagliaroli}\ \emph {et~al.}(2009)\citenamefont
  {Pagliaroli}, \citenamefont {Vissani}, \citenamefont {Costantini},\ and\
  \citenamefont {Ianni}}]{Pagliaroli2009-ci}%
  \BibitemOpen
  \bibfield  {author} {\bibinfo {author} {\bibfnamefont {G.}~\bibnamefont
  {Pagliaroli}}, \bibinfo {author} {\bibfnamefont {F.}~\bibnamefont {Vissani}},
  \bibinfo {author} {\bibfnamefont {M.~L.}\ \bibnamefont {Costantini}},\ and\
  \bibinfo {author} {\bibfnamefont {A.}~\bibnamefont {Ianni}},\ }\href@noop {}
  {\bibfield  {journal} {\bibinfo  {journal} {Astropart. Phys.}\ }\textbf
  {\bibinfo {volume} {31}},\ \bibinfo {pages} {163} (\bibinfo {year}
  {2009})}\BibitemShut {NoStop}%
\bibitem [{\citenamefont {O'Connor}\ and\ \citenamefont
  {Ott}(2013)}]{OConnor2013-at}%
  \BibitemOpen
  \bibfield  {author} {\bibinfo {author} {\bibfnamefont {E.}~\bibnamefont
  {O'Connor}}\ and\ \bibinfo {author} {\bibfnamefont {C.~D.}\ \bibnamefont
  {Ott}},\ }\href@noop {} {\bibfield  {journal} {\bibinfo  {journal}
  {Astrophys. J.}\ }\textbf {\bibinfo {volume} {762}},\ \bibinfo {pages} {126}
  (\bibinfo {year} {2013})}\BibitemShut {NoStop}%
\bibitem [{\citenamefont {Asimov}(1990)}]{Isaac}%
  \BibitemOpen
  \bibfield  {author} {\bibinfo {author} {\bibfnamefont {I.}~\bibnamefont
  {Asimov}},\ }\href@noop {} {\emph {\bibinfo {title} {Franchise}}}\ (\bibinfo
  {publisher} {Broadway Books, New York},\ \bibinfo {year} {1990})\BibitemShut
  {NoStop}%
\bibitem [{\citenamefont {Cowan}\ \emph {et~al.}(2011)\citenamefont {Cowan},
  \citenamefont {Cranmer}, \citenamefont {Gross},\ and\ \citenamefont
  {Vitells}}]{Cowan2011-iv}%
  \BibitemOpen
  \bibfield  {author} {\bibinfo {author} {\bibfnamefont {G.}~\bibnamefont
  {Cowan}}, \bibinfo {author} {\bibfnamefont {K.}~\bibnamefont {Cranmer}},
  \bibinfo {author} {\bibfnamefont {E.}~\bibnamefont {Gross}},\ and\ \bibinfo
  {author} {\bibfnamefont {O.}~\bibnamefont {Vitells}},\ }\href@noop {}
  {\bibfield  {journal} {\bibinfo  {journal} {Eur. Phys. J. C Part. Fields}\
  }\textbf {\bibinfo {volume} {71}},\ \bibinfo {pages} {1} (\bibinfo {year}
  {2011})}\BibitemShut {NoStop}%
\bibitem [{\citenamefont {{Borexino Collaboration}}\ \emph
  {et~al.}(2009)\citenamefont {{Borexino Collaboration}}, \citenamefont
  {Alimonti}, \citenamefont {Arpesella}, \citenamefont {Back},\ and\
  \citenamefont {\textit{et al}}}]{Alimonti2009}%
  \BibitemOpen
  \bibfield  {author} {\bibinfo {author} {\bibnamefont {{Borexino
  Collaboration}}}, \bibinfo {author} {\bibfnamefont {G.}~\bibnamefont
  {Alimonti}}, \bibinfo {author} {\bibfnamefont {C.}~\bibnamefont {Arpesella}},
  \bibinfo {author} {\bibfnamefont {H.}~\bibnamefont {Back}},\ and\ \bibinfo
  {author} {\bibfnamefont {M.~B.}\ \bibnamefont {\textit{et al}}},\ }\href
  {https://doi.org/10.1016/j.nima.2008.11.076} {\bibfield  {journal} {\bibinfo
  {journal} {Nuclear Instruments and Methods in Physics Research Section A:
  Accelerators, Spectrometers, Detectors and Associated Equipment}\ }\textbf
  {\bibinfo {volume} {600}},\ \bibinfo {pages} {568} (\bibinfo {year}
  {2009})}\BibitemShut {NoStop}%
\bibitem [{\citenamefont {{SNO+ Collaboration}}\ \emph
  {et~al.}(2021)\citenamefont {{SNO+ Collaboration}}, \citenamefont {Albanese},
  \citenamefont {Alves}, \citenamefont {Anderson}, \citenamefont {Andringa},\
  and\ \citenamefont {\textit{et al}}}]{Albanese2021}%
  \BibitemOpen
  \bibfield  {author} {\bibinfo {author} {\bibnamefont {{SNO+ Collaboration}}},
  \bibinfo {author} {\bibfnamefont {V.}~\bibnamefont {Albanese}}, \bibinfo
  {author} {\bibfnamefont {R.}~\bibnamefont {Alves}}, \bibinfo {author}
  {\bibfnamefont {M.}~\bibnamefont {Anderson}}, \bibinfo {author}
  {\bibfnamefont {S.}~\bibnamefont {Andringa}},\ and\ \bibinfo {author}
  {\bibfnamefont {L.~A.}\ \bibnamefont {\textit{et al}}},\ }\href
  {https://doi.org/10.1088/1748-0221/16/08/p08059} {\bibfield  {journal}
  {\bibinfo  {journal} {Journal of Instrumentation}\ }\textbf {\bibinfo
  {volume} {16}}\bibinfo  {number} { (08)},\ \bibinfo {pages}
  {P08059}}\BibitemShut {NoStop}%
\bibitem [{\citenamefont {Wurm}\ \emph {et~al.}(2012)\citenamefont {Wurm},
  \citenamefont {Beacom}, \citenamefont {Bezrukov}, \citenamefont {Bick},\ and\
  \citenamefont {\textit{et al}}}]{Wurm2012}%
  \BibitemOpen
\bibfield  {number} {  }\bibfield  {author} {\bibinfo {author} {\bibfnamefont
  {M.}~\bibnamefont {Wurm}}, \bibinfo {author} {\bibfnamefont {J.~F.}\
  \bibnamefont {Beacom}}, \bibinfo {author} {\bibfnamefont {L.~B.}\
  \bibnamefont {Bezrukov}}, \bibinfo {author} {\bibfnamefont {D.}~\bibnamefont
  {Bick}},\ and\ \bibinfo {author} {\bibfnamefont {J.~B.}\ \bibnamefont
  {\textit{et al}}},\ }\href
  {https://doi.org/10.1016/j.astropartphys.2012.02.011} {\bibfield  {journal}
  {\bibinfo  {journal} {Astroparticle Physics}\ }\textbf {\bibinfo {volume}
  {35}},\ \bibinfo {pages} {685} (\bibinfo {year} {2012})}\BibitemShut
  {NoStop}%
\bibitem [{\citenamefont {{KamLAND RCNS Group}}\ \emph
  {et~al.}(2004)\citenamefont {{KamLAND RCNS Group}}, \citenamefont {Suekane},
  \citenamefont {Iwamoto}, \citenamefont {Ogawa}, \citenamefont {Tajima},\ and\
  \citenamefont {Watanabe}}]{KamLAND}%
  \BibitemOpen
  \bibfield  {author} {\bibinfo {author} {\bibnamefont {{KamLAND RCNS Group}}},
  \bibinfo {author} {\bibfnamefont {F.}~\bibnamefont {Suekane}}, \bibinfo
  {author} {\bibfnamefont {T.}~\bibnamefont {Iwamoto}}, \bibinfo {author}
  {\bibfnamefont {H.}~\bibnamefont {Ogawa}}, \bibinfo {author} {\bibfnamefont
  {O.}~\bibnamefont {Tajima}},\ and\ \bibinfo {author} {\bibfnamefont
  {H.}~\bibnamefont {Watanabe}},\ }\href
  {https://doi.org/10.48550/ARXIV.PHYSICS/0404071} {\bibinfo {title} {An
  overview of the kamland 1-kiloton liquid scintillator}} (\bibinfo {year}
  {2004})\BibitemShut {NoStop}%
\bibitem [{Abe(2023)}]{Abe2023}%
  \BibitemOpen
  \ \textbf {\bibinfo {volume} {{130}}},\ \href
  {https://doi.org/{10.1103/physrevlett.130.051801}}
  {{10.1103/physrevlett.130.051801}} (\bibinfo {year} {{2023}})\BibitemShut
  {NoStop}%
\bibitem [{\citenamefont {Abe}\ \emph {et~al.}(2008)\citenamefont {Abe},
  \citenamefont {Ebihara}, \citenamefont {Enomoto}, \citenamefont {Furuno},\
  and\ \citenamefont {\textit{et al}}}]{Abe2008}%
  \BibitemOpen
  \bibfield  {author} {\bibinfo {author} {\bibfnamefont {S.}~\bibnamefont
  {Abe}}, \bibinfo {author} {\bibfnamefont {T.}~\bibnamefont {Ebihara}},
  \bibinfo {author} {\bibfnamefont {S.}~\bibnamefont {Enomoto}}, \bibinfo
  {author} {\bibfnamefont {K.}~\bibnamefont {Furuno}},\ and\ \bibinfo {author}
  {\bibfnamefont {Y.~G.}\ \bibnamefont {\textit{et al}}},\ }\bibfield
  {journal} {\bibinfo  {journal} {Physical Review Letters}\ }\textbf {\bibinfo
  {volume} {100}},\ \href {https://doi.org/10.1103/physrevlett.100.221803}
  {10.1103/physrevlett.100.221803} (\bibinfo {year} {2008})\BibitemShut
  {NoStop}%
\bibitem [{\citenamefont {Hunt}\ \emph {et~al.}(2016)\citenamefont {Hunt},
  \citenamefont {Bovy},\ and\ \citenamefont {Carlberg}}]{Hunt2016-ua}%
  \BibitemOpen
  \bibfield  {author} {\bibinfo {author} {\bibfnamefont {J.~A.~S.}\
  \bibnamefont {Hunt}}, \bibinfo {author} {\bibfnamefont {J.}~\bibnamefont
  {Bovy}},\ and\ \bibinfo {author} {\bibfnamefont {R.~G.}\ \bibnamefont
  {Carlberg}},\ }\href@noop {} {\bibfield  {journal} {\bibinfo  {journal}
  {Astrophys. J. Lett.}\ }\textbf {\bibinfo {volume} {832}},\ \bibinfo {pages}
  {L25} (\bibinfo {year} {2016})}\BibitemShut {NoStop}%
\bibitem [{\citenamefont {Adams}\ \emph
  {et~al.}(2013{\natexlab{b}})\citenamefont {Adams}, \citenamefont {Kochanek},
  \citenamefont {Beacom}, \citenamefont {Vagins},\ and\ \citenamefont
  {Stanek}}]{Adams2013-qa}%
  \BibitemOpen
  \bibfield  {author} {\bibinfo {author} {\bibfnamefont {S.~M.}\ \bibnamefont
  {Adams}}, \bibinfo {author} {\bibfnamefont {C.~S.}\ \bibnamefont {Kochanek}},
  \bibinfo {author} {\bibfnamefont {J.~F.}\ \bibnamefont {Beacom}}, \bibinfo
  {author} {\bibfnamefont {M.~R.}\ \bibnamefont {Vagins}},\ and\ \bibinfo
  {author} {\bibfnamefont {K.~Z.}\ \bibnamefont {Stanek}},\ }\href@noop {}
  {\bibfield  {journal} {\bibinfo  {journal} {Astrophys. J.}\ }\textbf
  {\bibinfo {volume} {778}},\ \bibinfo {pages} {164} (\bibinfo {year}
  {2013}{\natexlab{b}})}\BibitemShut {NoStop}%
\bibitem [{\citenamefont {Firestone}(2014)}]{Firestone2014-oi}%
  \BibitemOpen
  \bibfield  {author} {\bibinfo {author} {\bibfnamefont {R.~B.}\ \bibnamefont
  {Firestone}},\ }\href@noop {} {\bibfield  {journal} {\bibinfo  {journal}
  {Astrophys. J.}\ }\textbf {\bibinfo {volume} {789}},\ \bibinfo {pages} {29}
  (\bibinfo {year} {2014})}\BibitemShut {NoStop}%
\bibitem [{\citenamefont {Borriello}\ \emph {et~al.}(2012)\citenamefont
  {Borriello}, \citenamefont {Chakraborty}, \citenamefont {Mirizzi},
  \citenamefont {Serpico},\ and\ \citenamefont {Tamborra}}]{Borriello2012-me}%
  \BibitemOpen
  \bibfield  {author} {\bibinfo {author} {\bibfnamefont {E.}~\bibnamefont
  {Borriello}}, \bibinfo {author} {\bibfnamefont {S.}~\bibnamefont
  {Chakraborty}}, \bibinfo {author} {\bibfnamefont {A.}~\bibnamefont
  {Mirizzi}}, \bibinfo {author} {\bibfnamefont {P.~D.}\ \bibnamefont
  {Serpico}},\ and\ \bibinfo {author} {\bibfnamefont {I.}~\bibnamefont
  {Tamborra}},\ }\href@noop {} {\bibfield  {journal} {\bibinfo  {journal}
  {Phys. rev.}\ }\textbf {\bibinfo {volume} {86}} (\bibinfo {year}
  {2012})}\BibitemShut {NoStop}%
\bibitem [{\citenamefont {Liao}(2016)}]{Liao2016-eo}%
  \BibitemOpen
  \bibfield  {author} {\bibinfo {author} {\bibfnamefont {W.}~\bibnamefont
  {Liao}},\ }\href@noop {} {\bibfield  {journal} {\bibinfo  {journal} {Phys.
  Rev. D.}\ }\textbf {\bibinfo {volume} {94}} (\bibinfo {year}
  {2016})}\BibitemShut {NoStop}%
\bibitem [{\citenamefont {Walk}\ \emph {et~al.}(2018)\citenamefont {Walk},
  \citenamefont {Tamborra}, \citenamefont {Janka},\ and\ \citenamefont
  {Summa}}]{Walk2018-bk}%
  \BibitemOpen
  \bibfield  {author} {\bibinfo {author} {\bibfnamefont {L.}~\bibnamefont
  {Walk}}, \bibinfo {author} {\bibfnamefont {I.}~\bibnamefont {Tamborra}},
  \bibinfo {author} {\bibfnamefont {H.-T.}\ \bibnamefont {Janka}},\ and\
  \bibinfo {author} {\bibfnamefont {A.}~\bibnamefont {Summa}},\ }\href@noop {}
  {\bibfield  {journal} {\bibinfo  {journal} {Phys. Rev. D.}\ }\textbf
  {\bibinfo {volume} {98}} (\bibinfo {year} {2018})}\BibitemShut {NoStop}%
\bibitem [{\citenamefont {Wang}\ \emph {et~al.}(2021)\citenamefont {Wang},
  \citenamefont {Tseng}, \citenamefont {Gullin},\ and\ \citenamefont
  {O'Connor}}]{Wang2021-og}%
  \BibitemOpen
  \bibfield  {author} {\bibinfo {author} {\bibfnamefont {J.-S.}\ \bibnamefont
  {Wang}}, \bibinfo {author} {\bibfnamefont {J.}~\bibnamefont {Tseng}},
  \bibinfo {author} {\bibfnamefont {S.}~\bibnamefont {Gullin}},\ and\ \bibinfo
  {author} {\bibfnamefont {E.~P.}\ \bibnamefont {O'Connor}},\ }\href@noop {}
  {\bibfield  {journal} {\bibinfo  {journal} {Phys. Rev. D.}\ }\textbf
  {\bibinfo {volume} {104}} (\bibinfo {year} {2021})}\BibitemShut {NoStop}%
\bibitem [{\citenamefont {Gullin}\ \emph {et~al.}(2022)\citenamefont {Gullin},
  \citenamefont {O'Connor}, \citenamefont {Wang},\ and\ \citenamefont
  {Tseng}}]{Gullin2022-ez}%
  \BibitemOpen
  \bibfield  {author} {\bibinfo {author} {\bibfnamefont {S.}~\bibnamefont
  {Gullin}}, \bibinfo {author} {\bibfnamefont {E.~P.}\ \bibnamefont
  {O'Connor}}, \bibinfo {author} {\bibfnamefont {J.-S.}\ \bibnamefont {Wang}},\
  and\ \bibinfo {author} {\bibfnamefont {J.}~\bibnamefont {Tseng}},\
  }\href@noop {} {\bibfield  {journal} {\bibinfo  {journal} {Astrophys. J.}\
  }\textbf {\bibinfo {volume} {926}},\ \bibinfo {pages} {212} (\bibinfo {year}
  {2022})}\BibitemShut {NoStop}%
\bibitem [{\citenamefont {O'Connor}\ and\ \citenamefont
  {Ott}(2010)}]{OConnor2010-gv}%
  \BibitemOpen
  \bibfield  {author} {\bibinfo {author} {\bibfnamefont {E.}~\bibnamefont
  {O'Connor}}\ and\ \bibinfo {author} {\bibfnamefont {C.~D.}\ \bibnamefont
  {Ott}},\ }\href@noop {} {\bibfield  {journal} {\bibinfo  {journal} {Class.
  Quantum Gravity}\ }\textbf {\bibinfo {volume} {27}},\ \bibinfo {pages}
  {114103} (\bibinfo {year} {2010})}\BibitemShut {NoStop}%
\bibitem [{\citenamefont {O'Connor}(2015)}]{OConnor2015-ju}%
  \BibitemOpen
  \bibfield  {author} {\bibinfo {author} {\bibfnamefont {E.}~\bibnamefont
  {O'Connor}},\ }\href@noop {} {\bibfield  {journal} {\bibinfo  {journal}
  {Astrophys. J. Suppl. Ser.}\ }\textbf {\bibinfo {volume} {219}},\ \bibinfo
  {pages} {24} (\bibinfo {year} {2015})}\BibitemShut {NoStop}%
\bibitem [{\citenamefont {{DUNE Collaboration}}\ and\ \citenamefont
  {Acciarri}(2015)}]{DUNE_Collaboration2015-sb}%
  \BibitemOpen
  \bibfield  {author} {\bibinfo {author} {\bibnamefont {{DUNE Collaboration}}}\
  and\ \bibinfo {author} {\bibfnamefont {R.}~\bibnamefont {Acciarri}},\
  }\href@noop {} {\  (\bibinfo {year} {2015})},\ \Eprint
  {https://arxiv.org/abs/1512.06148} {arXiv:1512.06148 [physics.ins-det]}
  \BibitemShut {NoStop}%
\bibitem [{\citenamefont {Kharusi}\ \emph {et~al.}(2020)\citenamefont
  {Kharusi}, \citenamefont {BenZvi}, \citenamefont {Bobowski}, \citenamefont
  {Bonivento},\ and\ \citenamefont {Brdar}}]{Kharusi2020-bk}%
  \BibitemOpen
  \bibfield  {author} {\bibinfo {author} {\bibfnamefont {S.~A.}\ \bibnamefont
  {Kharusi}}, \bibinfo {author} {\bibfnamefont {S.~Y.}\ \bibnamefont {BenZvi}},
  \bibinfo {author} {\bibfnamefont {J.~S.}\ \bibnamefont {Bobowski}}, \bibinfo
  {author} {\bibfnamefont {W.}~\bibnamefont {Bonivento}},\ and\ \bibinfo
  {author} {\bibfnamefont {V.~t.}\ \bibnamefont {Brdar}},\ }\href@noop {} {\
  (\bibinfo {year} {2020})},\ \Eprint {https://arxiv.org/abs/2011.00035}
  {arXiv:2011.00035 [astro-ph.HE]} \BibitemShut {NoStop}%
\bibitem [{\citenamefont {Al~Kharusi}\ \emph {et~al.}(2021)\citenamefont
  {Al~Kharusi}, \citenamefont {BenZvi}, \citenamefont {Bobowski}, \citenamefont
  {Bonivento},\ and\ \citenamefont {Brdar}}]{Al_Kharusi2021-uf}%
  \BibitemOpen
  \bibfield  {author} {\bibinfo {author} {\bibfnamefont {S.}~\bibnamefont
  {Al~Kharusi}}, \bibinfo {author} {\bibfnamefont {S.~Y.}\ \bibnamefont
  {BenZvi}}, \bibinfo {author} {\bibfnamefont {J.~S.}\ \bibnamefont
  {Bobowski}}, \bibinfo {author} {\bibfnamefont {W.}~\bibnamefont
  {Bonivento}},\ and\ \bibinfo {author} {\bibfnamefont {V.~t.}\ \bibnamefont
  {Brdar}},\ }\href@noop {} {\bibfield  {journal} {\bibinfo  {journal} {New J.
  Phys.}\ }\textbf {\bibinfo {volume} {23}},\ \bibinfo {pages} {031201}
  (\bibinfo {year} {2021})}\BibitemShut {NoStop}%
\bibitem [{\citenamefont {Abud}(2021)}]{Abud2021-oi}%
  \BibitemOpen
  \bibfield  {author} {\bibinfo {author} {\bibfnamefont {A.~A.}\ \bibnamefont
  {Abud}},\ }\href@noop {} {\  (\bibinfo {year} {2021})}\BibitemShut {NoStop}%
\bibitem [{\citenamefont {Yang}\ \emph {et~al.}(2018)\citenamefont {Yang},
  \citenamefont {Yu}, \citenamefont {Cao}, \citenamefont {Sun}, \citenamefont
  {Yu},\ and\ \citenamefont {An}}]{Yang2018-yx}%
  \BibitemOpen
  \bibfield  {author} {\bibinfo {author} {\bibfnamefont {M.~S.}\ \bibnamefont
  {Yang}}, \bibinfo {author} {\bibfnamefont {Z.~Y.}\ \bibnamefont {Yu}},
  \bibinfo {author} {\bibfnamefont {J.}~\bibnamefont {Cao}}, \bibinfo {author}
  {\bibfnamefont {X.~L.}\ \bibnamefont {Sun}}, \bibinfo {author} {\bibfnamefont
  {B.~X.}\ \bibnamefont {Yu}},\ and\ \bibinfo {author} {\bibfnamefont {G.~P.}\
  \bibnamefont {An}},\ }\href@noop {} {\  (\bibinfo {year} {2018})}\BibitemShut
  {NoStop}%
\end{thebibliography}%

\appendix
\section{\label{sec:level13}{Other Channels}}

A complimentary study for electron neutrinos, $\Pnue$, could be performed in the Deep Underground Neutrino Experiment (DUNE), is a liquid argon time-projection chamber with a fiducial mass of 40\unit{\kilo\tonne} \cite{Abud2021-oi}. The main detection channel on argon is the charged-current (CC) reaction with $\Pnue$,
\begin{equation*}
\Pnue + {}^{40}\textrm{Ar} \rightarrow \Pelectron + {}^{40}\textrm{K}^{*}.
\end{equation*}

This $\Pnue$ sensitivity is unique to DUNE, and will allow a detailed look at the neutronization burst of the CCSN neutrino curve \cite{DUNE_Collaboration2015-sb}. The complimentary analysis in DUNE could use the neutronization burst as a similarly clear signature as black hole formation to carry out a ToF analysis.

Although it is less suitable for studies involving ToF effects, JUNO presents a unique opportunity to detect $\nu_x$ from CCSN. Proton-elastic scattering $\nu + p \rightarrow \nu + p$ is a neutral-current interaction only available to low-threshold scintillator detectors \cite{Yang2018-yx}, giving JUNO another perspective on the engine of core-collapse.

\end{document}